\documentclass{jfm}
\usepackage{graphicx}
\usepackage{epstopdf, epsfig}
\usepackage{amsmath}
\usepackage{adjustbox}
\usepackage{upgreek}
\usepackage{comment}
\usepackage{booktabs}
\usepackage{amssymb}
\usepackage{makecell}
\usepackage{color}
\usepackage{rotating}
\usepackage{pdflscape}

\shorttitle{Wind-farm wake behaviour in CNBLs}
\shortauthor{L. Lanzilao and J. Meyers}

\title{Wind-farm wake recovery mechanisms in conventionally neutral boundary layers}
\author{L. Lanzilao\aff{1} \corresp{\email{luca.lanzilao@kuleuven.be}}
	\and J. Meyers\aff{1}}

\affiliation{\aff{1}KU Leuven, Department of Mechanical Engineering, Celestijnenlaan 300 – box 2421, B-3001 Leuven, Belgium} 

\begin{document}

\maketitle

\begin{abstract}
Synthetic-aperture radar images and mesoscale model results show that wind-farm wakes behave very differently than single-turbine wakes, e.g. with wakes that seemingly narrow and do not disperse over long distances. In the current work, we aim at better understanding the physical mechanisms that govern wind-farm wake behaviour and recovery. Hence, we study the wake properties of a $1.6$~GW wind-farm operating in conventionally neutral boundary layers with four capping-inversion heights, i.e. $203$, $319$, $507$ and $1001$~m. In shallow boundary layers, we find strong flow decelerations which reduce the Coriolis force magnitude, leading to an anticlockwise wake deflection in the Northern Hemisphere. In deep boundary layers, the vertical turbulent entrainment of momentum adds clockwise-turning flow from aloft into the wake region, leading to a faster recovery rate and a clockwise wake deflection. To estimate the wake properties, we develop a simple model that fits the velocity magnitude profiles along the spanwise direction. Based on this, we observe that the wake narrows along the downstream direction in all cases. Further, a detailed momentum budget analysis shows that the wake is mostly replenished by turbulent vertical entrainment in deep boundary layers. In shallow boundary layers, the capping inversion limits vertical motions and wakes are mostly replenished by mean flow entrainment in the spanwise direction. Moreover, in these cases, we observe a counterclockwise flow rotation near the left edge of the wake, which persists at each location downstream of the farm, giving rise to local strong streamwise velocity gradients along the spanwise direction. 
\end{abstract}


\section{Introduction}\label{sec:intoduction}
\begin{figure}
	\centering
	\includegraphics[width=0.75\textwidth]{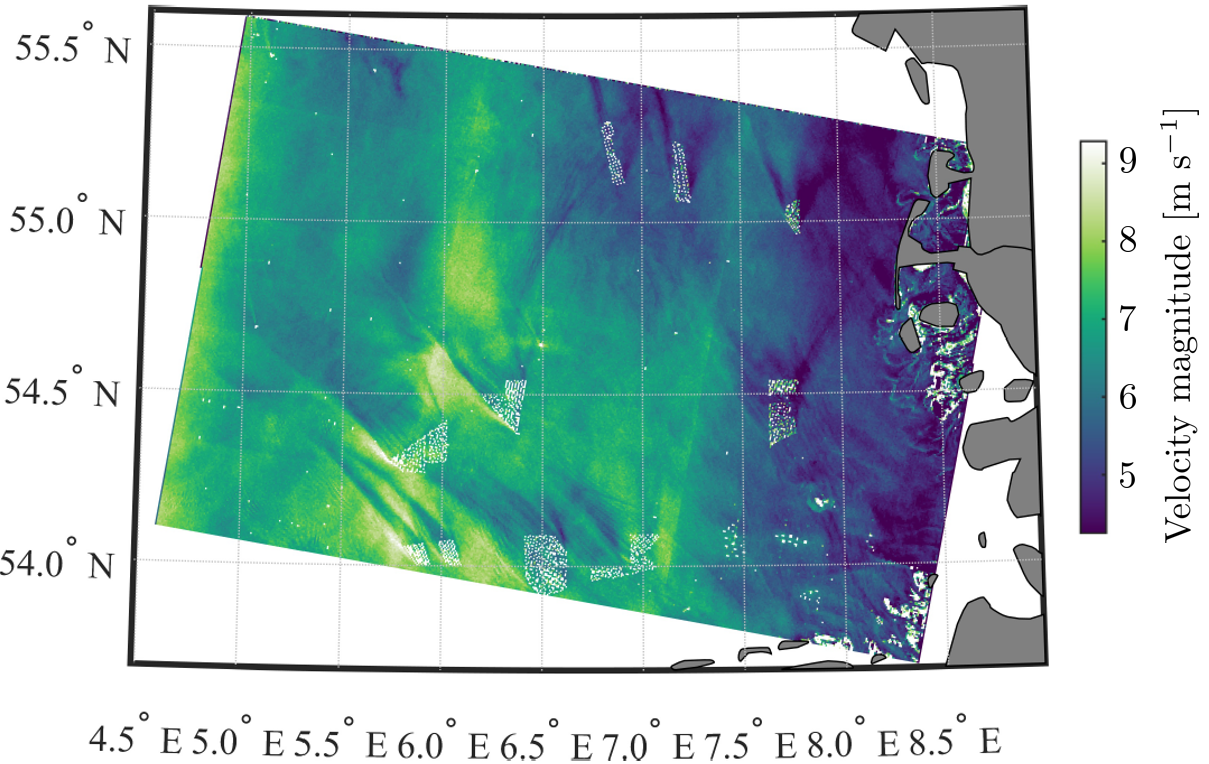}
	\caption{Velocity magnitude at 10 m above the sea surface observed over the German Bight by the Sentinel-1A satellite on 17-04-2022. Wind turbines are visible as white dots. Figure adapted from \cite{Finseraas2023}.}
	\label{fig:sar_wake}
\end{figure}

The wind energy sector has established itself as the primary non-hydro renewable energy technology, with a total installation of 117 GW in 2023, representing a 50\% year-on-year increase from 2022 \citep{GWEC2024}. Although this rapid growth contributes to the reduction of billion tons of C02 emissions each year, it also poses some challenges. In fact, favourable offshore regions with shallow water depths, out of fishing zones and maritime shipping routes, are limited, and as a result wind farms are clustered together. For instance wind farms located in the North Sea are often less than $50$~km away from their nearest neighbour \citep{Finseraas2023}. Although  capacity factors are typically higher offshore, such limited distances between farms can significantly reduce the annual energy production \citep{Pryor2001,Baas2022}. Wake shadowing effects are clearly visible in Figure~\ref{fig:sar_wake}, which illustrates an example of the reconstructed velocity magnitude from synthetic-aperture radar (SAR) images over the German Bight area. Here, we observe wakes longer than 70 km, which seems to narrow and slowly disperse. The length of these wakes could also lead to conflicts, since numerous neighbouring wind farms are operated by different owners \citep{Kenis2023,Finseraas2023}. As a result, the understanding of wind-farm wake strength and deflection together with their impact on neighbouring farms constitutes one of the major challenges today. In the current work we contribute towards this direction, aiming at better understanding the wind-farm wake behaviour and recovery mechanisms in conventionally neutral boundary layers (CNBLs).

The presence of offshore wind-farm wakes have been detected in SAR images already in 2005, when \cite{Christiansen2005} noticed a 2\% velocity reduction $20$~km downwind of the Horns Rev farm, a small wind-farm compared to today's standards. A similar behaviour was later noticed by \cite{Hasager2015b}, who observed wakes longer than $50$~km for wind farms located in the Southern part of the North Sea. Offshore wind-farm wakes have also been detected by using dual-Doppler radar techniques \citep{Ahsbash2020,Abraham2024}. For instance, \cite{Nygaard2018} observed strong velocity deficits 17 km downwind the Westermost Rough farm, a relatively small wind farm. More recently, \cite{Djath2018} investigated wakes generated by cluster of wind-farms using SAR images, noticing velocity deficit up to 70 km downwind. Moreover, they also observed longer wakes in stable atmospheric conditions, mostly due to the lower level of turbulence intensity which reduces flow mixing and consequently the wake recovery rate \citep{Emeis2016}. Similar conclusions were drawn by \cite{Lundquist2019} and \cite{Schneemann2020}, who noticed that, in stably stratified atmospheric conditions, wakes can extend more than $50$~km downwind. Wake shadowing effects can be seen also in supervisory control and data acquisition (SCADA) data \citep{Ahsbahs2018}. For instance, \cite{Nygaard2014} compared the power output of the Nysted farm before and after the construction of the R\o{}dsand II farm, often located upwind, observing wind-turbine efficiency reductions up to $30\%$. A further example of the negative impact of wind-farm wakes is given in the work of \cite{Platis2018}, who captured wind-farm wakes by in situ measurements taken by a specially equipped aircraft. Their results confirmed the presence of wakes longer than $40$ km with maximum velocity deficits of $40\%$ with respect to the background wind speed.

The observations mentioned above are supported by several numerical studies. The large computational domain typically required by this type of studies favoured the use of low-fidelity analytical wake models \citep{Bastankhan2014,Gocmen2016,Blondel2020,Lanzilao2021b}. \cite{Nygaard2020} investigated the influence of the Humber Gateway wind farm on the Westermost Rough with an analytical flow model, finding that the wake impact on the front row turbines is up to $30\%$ despite the $15$~km distance. \cite{Munters2022} adopted a similar modelling approach, finding an annual energy production (AEP) loss due to inter-cluster wakes up to $0.8\%$. However, these analytical models have large uncertainties and are known to over-predict the wind-farm wake recovery rate compared to LES \citep{Stieren2021b,Maas2023}. In an attempt to improve on these results, \cite{Bastankhah2024} recently developed an analytical fast-running physics-based wake model suitable for a computationally inexpensive prediction of wind-farm wake strength and deflection, finding very good agreement with LES results. However, their model is only suitable for spanwise-infinite wind farms. 

Several studies have also been performed using higher fidelity models. \cite{Vanderlaan2015b} and \cite{Vanderlaan2017b} used Reynolds-Averaged Navier-Stokes (RANS) simulations to investigate the effects of the Coriolis force on the wind-farm wake development, highlighting the importance of predicting the wake deflection. They found that the Coriolis force has two opposing effects on the wake deflection. Specifically, the direct effect induces an anticlockwise flow rotation of the wake in the Northern Hemisphere, while the indirect effect, mediated by wind veer in the background flow, causes a clockwise rotation. The same mechanism has been observed and discussed in single-turbine simulations by \cite{Heck2024} and in spanwise-infinite wind-farm simulations by \cite{Allaerts2017} and \cite{Bastankhah2024}, respectively. In an attempt to reduce computational costs, \cite{Vanderlaan2023} developed a new RANS-based wind-farm model for the specific application of investigating wind-farm wake effects on downstream clusters, which compared well against measurements. Although computationally expensive, several LES studies have also been performed. For instance, \cite{Baas2022} examined the potential impact of surrounding wind farms on the production of the planned 4~GW IJmuiden~Ver wind farm, finding production deficits of $4\%$ on a yearly basis. Later, \cite{Maas2022} used a LES framework for investigating the wind-farm wake development of spanwise-infinite small and large wind farms. They concluded that the wind-farm size has no impact on the decay of the turbulence intensity. However, the flow physics and wake recovery mechanisms differ, with the large farm triggering an inertial wave which causes the flow to accelerate far downstream. \cite{Stieren2022} also performed an LES of two identical wind farms separate by $5$ to $15$ km. They found that the power production of the first-row turbine of the downstream wind farm is reduced up to $33\%$. However, they observed that the upstream farm enhances vertical flow mixing, which results in a faster recovery rate for the waked wind farm. We would like to mention that many studies on farm-farm interactions have also been performed with mesoscale models, such as the Weather Research and Forecasting (WRF) model and the COSMO-CLM model \citep{Pryor2019,Pryor2021,Pryor2022,Pryor2024,Akhtar2021,Fischereit2022,Fischereit2022b,Borgers2024,Rosencrans2024}. These studies typically focus on the capacity factor losses generated by wake shadowing effects, therefore often considering multi-year long scenarios. However, due to the coarser grid resolution compared to LES, these studies do not discuss in details the dominant wake recovery mechanisms.

The majority of the studies mentioned above include the effects of both wind shear and veer on the wake development but they often make use of neutral atmospheric boundary layers topped with an artificial rigid lid, where thermal stratifications are not taken into account. However, \cite{Allaerts2017,Allaerts2017b,Allaerts2019,Bleeg2018,Bleeg2022,Lanzilao2022,Maas2022b,Lanzilao2024,Stipa2024} among others, have shown that the thermal stratification within and above the ABL have strong impact on the wind-farm performance and therefore it is expected to also influence the wake recovery mechanisms. \cite{Lanzilao2024} studied in details the effects of thermal stratification above the ABL on the wind-farm performance, but they did not focus on the wake behaviour. \cite{Maas2022b} were among the first to study farm--farm interactions over the German Bight area using a LES framework with various type of thermal stratifications. They observed wakes longer than $100$~km in stable boundary layers and they show that large-scale wind farms trigger gravity waves which influence the operation of smaller wind farms nearby. 

To date, there is a lack of understanding about the main mechanisms that drive the wind-farm wake recovery process. Moreover, SAR images and mesoscale model simulations seem to show that wind-farm wakes narrow and in general behave differently than single-turbine wakes or classic axisymmetric wakes in general (e.g see Figure \ref{fig:sar_wake}). In the current article we aim to fill this gap by bringing new physical insights into the wake behaviour of large-scale wind farms operating in CNBLs with capping-inversion heights varying from $200$ to $1000$~m, using an LES framework. The article is structured as follows. The simulation set-up is elaborated in Section~\ref{sec:methodology}. Thereafter, Section~\ref{sec:bl_initialization} discusses the boundary-layer initialization. Next, the wind-farm wake development and its recovery mechanisms are investigated in Section~\ref{sec:sensitivity_atm}, where we discuss the wake strength and deflection as a function of the inversion-layer height. Moreover, we develop a simple model that fits the velocity magnitude along the spanwise direction, which is fast and efficient for estimating wake properties such as the wake edges and center. Thereafter, we perform a momentum budget analysis, where we discuss in details the dominant terms that replenish the wind-farm wake, and we investigate the flow behaviour within the wake region and at its sides. Finally, conclusions are drawn in Section~\ref{sec:conclusions}.

\section{Methodology}\label{sec:methodology}
In the current study, we focus on the wake behaviour of a 1.6 GW wind farm operating in four CNBLs with capping-inversion heights of $203$, $319$, $507$ and $1001$ m. The filtered Navier--Stokes equations with Boussinesq approximation coupled with a transport equation for the potential temperature are used to investigate the flow behaviour in and around the wind farm \citep{Allaerts2017,Lanzilao2022,Lanzilao2024}. The equations are solved using the SP-Wind solver, an in-house software developed over the past 15 years at KU Leuven \citep{Meyers2007,Calaf2010,Goit2015,Allaerts2017,Munters2018,Allaerts2017b,Lanzilao2022,Lanzilao2022b,Lanzilao2024}. The flow solver is described in Section \ref{sec:flow_solver}. Next, we provide a summary of the numerical set-up and boundary conditions in Section~\ref{sec:numerical_setup} while the wind-farm layout and atmospheric states considered are described in Sections~\ref{sec:farm_setup} and \ref{sec:atm_state}, respectively.

\subsection{Flow solver}\label{sec:flow_solver}
The SP-Wind solver structure adopted here is mainly based on the version developed and used in \cite{Allaerts2017} and \cite{Lanzilao2022,Lanzilao2022b,Lanzilao2024}. The equations are advanced in time using a classic fourth-order Runge--Kutta scheme with a time step based on a Courant--Friedrichs--Lewy number of $0.4$. The streamwise~($x$) and spanwise ($y$) directions are discretized with a Fourier pseudo-spectral method. This implies that all linear terms are discretized in the spectral domain while non-linear operations are computed in the physical domain, reducing the cost of convolutions from quadratic to log-linear \citep{Fornberg1996}. Further, the 3/2 dealiasing technique is adopted to avoid aliasing errors \citep{Canuto1988}. For the vertical dimension~($z$), an energy-preserving fourth-order finite difference scheme is adopted~\citep{Verstappen2003}. Continuity is enforced by solving the Poisson equation during every stage of the Runge--Kutta scheme. The effects of subgrid-scale motions on the resolved flow are taken into account with the stability-dependent Smagorinsky model proposed by \cite{Stevens2000} with Smagorinsky coefficient set to $C_s=0.14$, similarly to previous studies performed with SP-Wind \citep{Goit2015,Allaerts2017,Munters2018}. The constant $C_s$ is damped near the wall by using the damping function proposed by \cite{Mason1992}. The turbines are modelled using a non-rotating actuator disk model (ADM) \cite{Meyers2010,Calaf2010}. We refer to \cite{DelportPhD} for more details on the discretization of the continuity and momentum equations while the implementation of the thermodynamic equation and sub-grid scale model are explained in detail in \cite{AllaertsPhD}.

\subsection{Numerical set-up}\label{sec:numerical_setup}
\begin{figure}
	\centering
	\includegraphics[width=0.7\textwidth]{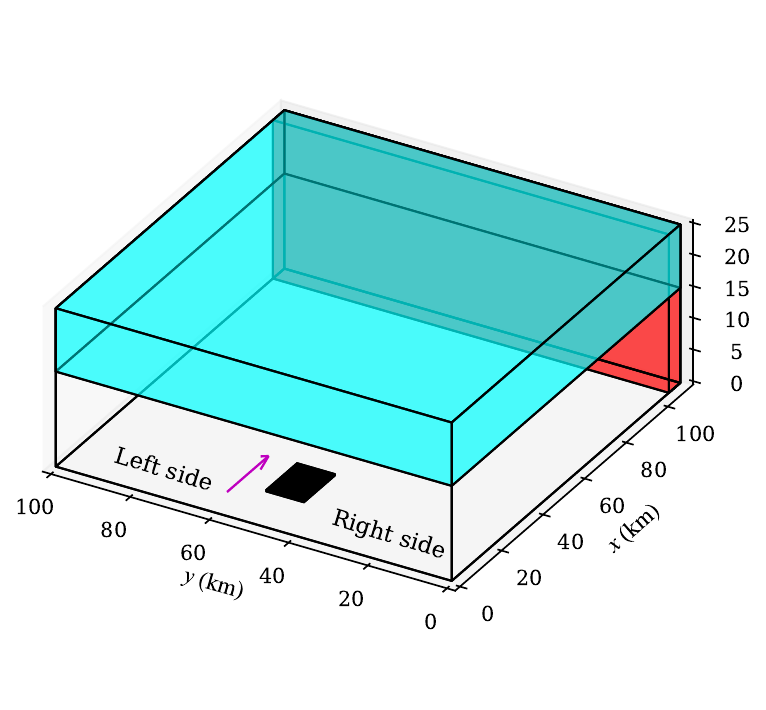}
	\caption{Sketch of the main domain adopted in this study. The Rayleigh damping layer and fringe region are represented with a cyan and red color, respectively, while the wind farm is denoted with a black rectangle. The pink arrow shows the flow direction at hub height. The left and right sides of the farm are named based on the direction of the flow, i.e. while looking downstream.}
	\label{fig:domain_sketch}
\end{figure}

The flow solver makes use of two numerical domains concurrently marched in time, i.e. the precursor and main domains. The precursor domain does not contain turbines and is only used for generating a turbulent fully developed statistically steady flow. The latter is then adopted for driving the simulation in the main domain. Similarly to \cite{Allaerts2017,Allaerts2017b} and \cite{Lanzilao2024}, we fix the precursor domain length and width to $L_x^p=L_y^p=10$ km, with $L_z^p=3$ km. The wind farm is located in the main domain, which should be sufficiently large to avoid spurious effects introduced by the domain boundaries. In Appendix \ref{app:domain_sensitivity}, we analyze in detail the effects of the domain width on the wake evolution. Based on this analysis, we select a domain size of $L_x \times L_y = 110 \times 100$ km$^2$. Following previous studies, we fix the main domain height to $L_z=25$ km \citep{Allaerts2017,Allaerts2017b,Lanzilao2022,Lanzilao2022b,Lanzilao2024}. Such a vertical extent allows gravity waves to decay and radiate energy outward, therefore minimizing reflectivity. A sketch of the main domain is reported in Figure \ref{fig:domain_sketch}. We note that the precursor domain width and height should match those of the main domain when they are run concurrently. Therefore, after the precursor spin-up phase is completed, we adopt the technique described in \cite{Sanchez2023} and \cite{Lanzilao2024} to extend the precursor flow fields in the $y$ direction from $10$ to $100$ km and in the $z$ direction from $3$ to $25$ km.

For the grid resolution, we fix $\Delta x = 62.5$ m and $\Delta y = 43.48$ m in the streamwise and spanwise direction, respectively. This leads to $N_x=1760$ and $N_y~=~2300$ grid points for the main domain and to $N_x^p=160$ and $N_y^p=230$ points for the precursor domain. We note that the resolution adopted in this work is twice as coarse in both the streamwise and spanwise directions as the one adopted in \cite{Lanzilao2022,Lanzilao2022b,Lanzilao2024}. This choice was dictated by the need of a sufficiently long and wide domain for studying the wake behaviour without spurious effects from the domain boundaries. We show in Appendix \ref{app:domain_sensitivity} that this change in grid resolution has very limited impact on the results. In the vertical direction, we adopt a stretched grid which corresponds to the one used in \cite{Lanzilao2022,Lanzilao2022b,Lanzilao2024}, i.e. with a resolution of $5$ m within the first $1.5$ km and stretched above, for a total of $490$ grid points. The combination of precursor and main domains leads to a total of roughly $8.65 \times 10^9$ degrees of freedom (DOF). We note that this number is evaluated as the product between the number of grid cells and the number of variables (i.e. $u$, $v$, $w$ and $\theta$, which denote the streamwise, spanwise and vertical velocity and potential-temperature field).

\begin{figure}
	\centering
	\includegraphics[width=1\textwidth]{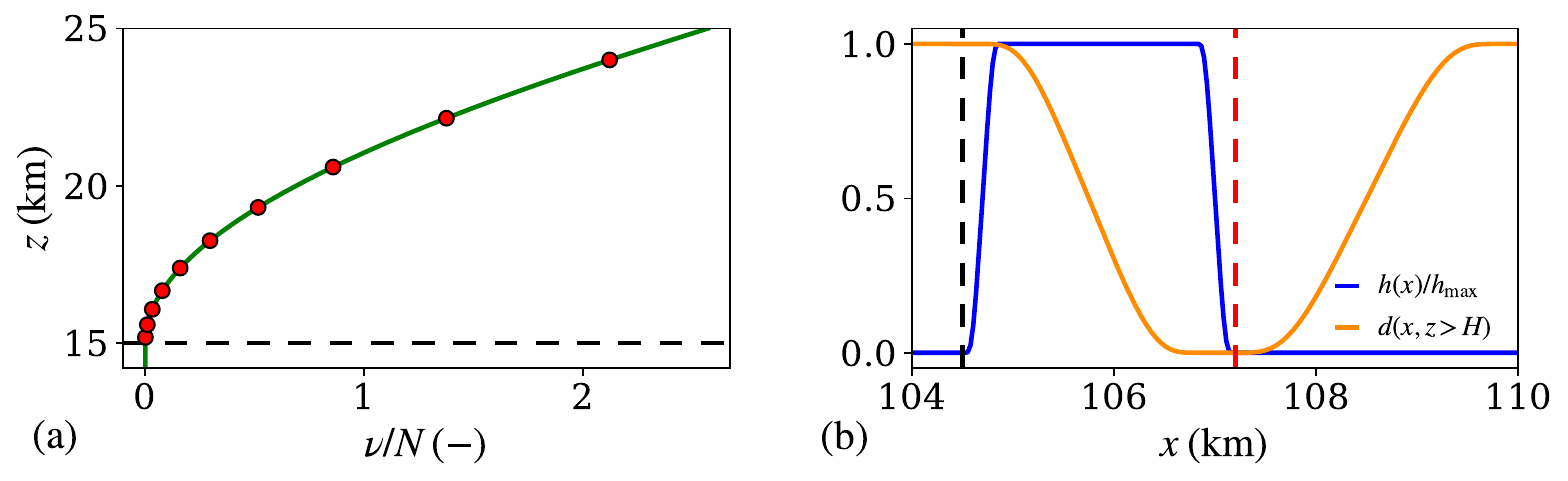}
	\caption{(a) Rayleigh function obtained with $\nu^\mathrm{ra}=5.15$ and $s^\mathrm{ra}=3$ values, normalized with the Brunt--V\"{a}is\"{a}l\"{a} frequency $N$. The black horizontal dashed line denote the start of the RDL. The parameter $\nu^\mathrm{ra}$ and $s^\mathrm{ra}$ control the magnitude and the gradient along the $z$ direction of the RDL function, respectively. The RDL is $10$ km long but only $10$ grid cells along the vertical direction are used in this region. The center of each grid cell is denoted with a red dot. (b) Fringe and damping functions. The total length of the buffer region (which we denote as fringe region) is of $L_x^\mathrm{fr}=5.5$ km. The support of the fringe function extends from $x_s^h=L_x-L_x^\mathrm{fr}$ to $x_e^h=L_x-2.8$ km while the upward and downward function slopes measure $\delta_s^h=\delta_e^h= 0.4$ km. The strength of the fringe forcing is set to $h_\mathrm{max}=0.3$~s$^{-1}$. The damping function starts to influence the flow at $x_s^d=x_s^h$ and is effective up to $x_e^d=L_x$. The downward and upward function slopes measure $\delta_s^d= 2.5$ km and $\delta_e^d= 3$ km, respectively. Above the capping-inversion height, the damping function assumes value of $1$ everywhere. We note that this setup corresponds to the one previously used by \cite{Lanzilao2024}. Finally, the black vertical dashed line denotes the start of the fringe region while the red vertical dashed line marks the end of the fringe forcing. For more details, we refer the reader to \cite{Lanzilao2022b,Lanzilao2024}. }
	\label{fig:buffer_regions}
\end{figure}

The top boundary conditions are a zero stress condition for the horizontal velocity, a zero vertical velocity, and a fixed potential temperature. Moreover, to minimize gravity-wave reflection, we adopt the Rayleigh damping layer (RDL) which extends from $15$ to $25$ km. For the lower boundary condition, we employ a classic wall stress formulation based on the Monin-Obukhov similarity theory for neutral boundary layers \citep{Moeng1984,Allaerts2015}. To avoid periodicity in the streamwise direction, we adopt the wave-free fringe-region technique developed by \cite{Lanzilao2022b}, which uses an additional convection-damping region to limit spurious gravity-wave effects. The fringe region is located at the end of the main domain, starting at $x=104.5$ km and is $5.5$ km long. The location and dimension of the buffer regions are illustrated in Figure \ref{fig:domain_sketch} while for more information about the fringe and damping functions together with the RDL setup, we refer the reader to Figure \ref{fig:buffer_regions}. 

\subsection{Wind-farm set-up}\label{sec:farm_setup}
The wind-farm layout and wind turbine choice correspond to the ones previously used by \cite{Lanzilao2022,Lanzilao2022b,Lanzilao2024}. Hence, the farm has $16$ rows and $10$ columns, for a total of $160$ turbines. We adopt the 10 MW IEA offshore turbine, which has a rotor diameter $D$ of $198$ m and a hub-height $z_h$ of $119$ m, leading to a tip height $z_\mathrm{tip}$ of 218 m \citep{Bortolotti2019}. Moreover, we fix the thrust coefficient $C_T'=1.94$, which corresponds to a $C_T$ of $0.88$. The turbines are arranged in a staggered layout with respect to the main wind direction with streamwise and spanwise spacings set to $S_x=S_y=5D$, resulting into a farm length and width of $L_x^f=14.85$ and $L_y^f=9.4$ km, respectively. Moreover, a simple yaw controller is implemented to keep the turbine-rotor disks perpendicular to the incident wind flow measured one rotor diameter upstream. The first row of turbines is located $15$ km from the inflow while the distance between the last row and the starting of the fringe region amounts to roughly $75$ km. We selected this configuration to enable the full development of the wind-farm wake, thereby facilitating a detailed examination of the mechanisms involved in the wake recovery. Moreover, the ratio $L_y/L_y^f$ measures $10.64$, which is well in line with the guidelines proposed by \cite{Lanzilao2024}. The main domain sketch reported in Figure \ref{fig:domain_sketch} further illustrates the wind-farm location and its size relative to the computational domain.

\subsection{Atmospheric state}\label{sec:atm_state}
The initial atmospheric states are chosen based on the analysis performed in \cite{Lanzilao2024}. Hence, we fix the geostrophic wind to $10$ m s$^{-1}$, so that all turbines operate below their rated power and in a region where the thrust curve typically shows a rather constant thrust-coefficient value. In regard to the capping-inversion height $H$, we initialize its value to $150$, $300$, $500$ and $1000$ m. This allows us to explore farm operations in shallow and deep boundary layers. The capping inversion strength $\Delta \theta$ is set to $5$ K while we fix the free-atmosphere lapse rate $\Gamma$ to $4$ K km$^{-1}$. Further, the ground temperature and the capping-inversion thickness are fixed to $\theta_0=288.15$~K and $\Delta H = 100$ m for all simulations. Finally, we select a latitude of $\phi=51.6^\circ$, which leads to a Coriolis frequency of $f_c=1.14 \times 10^{-4}$~s$^{-1}$ and a surface roughness of $z_0 = 1 \times 10^{-4}$~m. This value represents calm sea conditions and enters in the range of values observed over the North Sea, and more generally offshore \citep{Taylor2000,Allaerts2017,Lanzilao2022,Kirby2022}. Since the only changing parameter among the four cases considered is the capping inversion height, we denote these cases as H150, H300, H500 and H1000. We note that during the spin-up phase of the precursor, the value of the capping-inversion height slightly grow -- see Section \ref{sec:cnbls_spinup}.

In the remainder of the text, the state variables will be accompanied by a bar in case of time averages. For the horizontal averages along the full streamwise and spanwise directions, we use the angular brackets $\langle \cdot \rangle$ while the notations $\langle \cdot \rangle_{\!f}$ and $\langle \cdot \rangle_{\!w}$ is used to represent spanwise averages along the farm and wake width, respectively. Finally, we note that the RDL and fringe region will be left out of the figures in the remainder of the text.

\section{Boundary-layer initialization}\label{sec:bl_initialization}
In this section, we summarize the methodology applied to spin-up the precursor and wind-farm simulations. We note that the methodology follows the one adopted by \cite{Lanzilao2024}. Therefore we refer the reader to their work for more details. 

\subsection{Generation of a fully developed turbulent flow field}\label{sec:cnbls_spinup}
The initial vertical potential-temperature profiles are generated giving the $H$, $\Delta H$, $\Delta \theta$ and $\Gamma$ values as input to the \cite{Rampanelli2004} model. For the initial velocity profile, we use a constant geostrophic wind above the capping inversion. Within the ABL, we use the \cite{Zilitinkevich1989} model with friction velocity $u_\ast = 0.26$~m s$^{-1}$, which is in the range of values observed by \cite{Brost1982}. The velocity profiles below the capping inversion are then combined with the laminar profile in the free atmosphere following the method proposed by \cite{Allaerts2015}. 

\begin{figure}
	\centering
	\includegraphics[width=1.\textwidth]{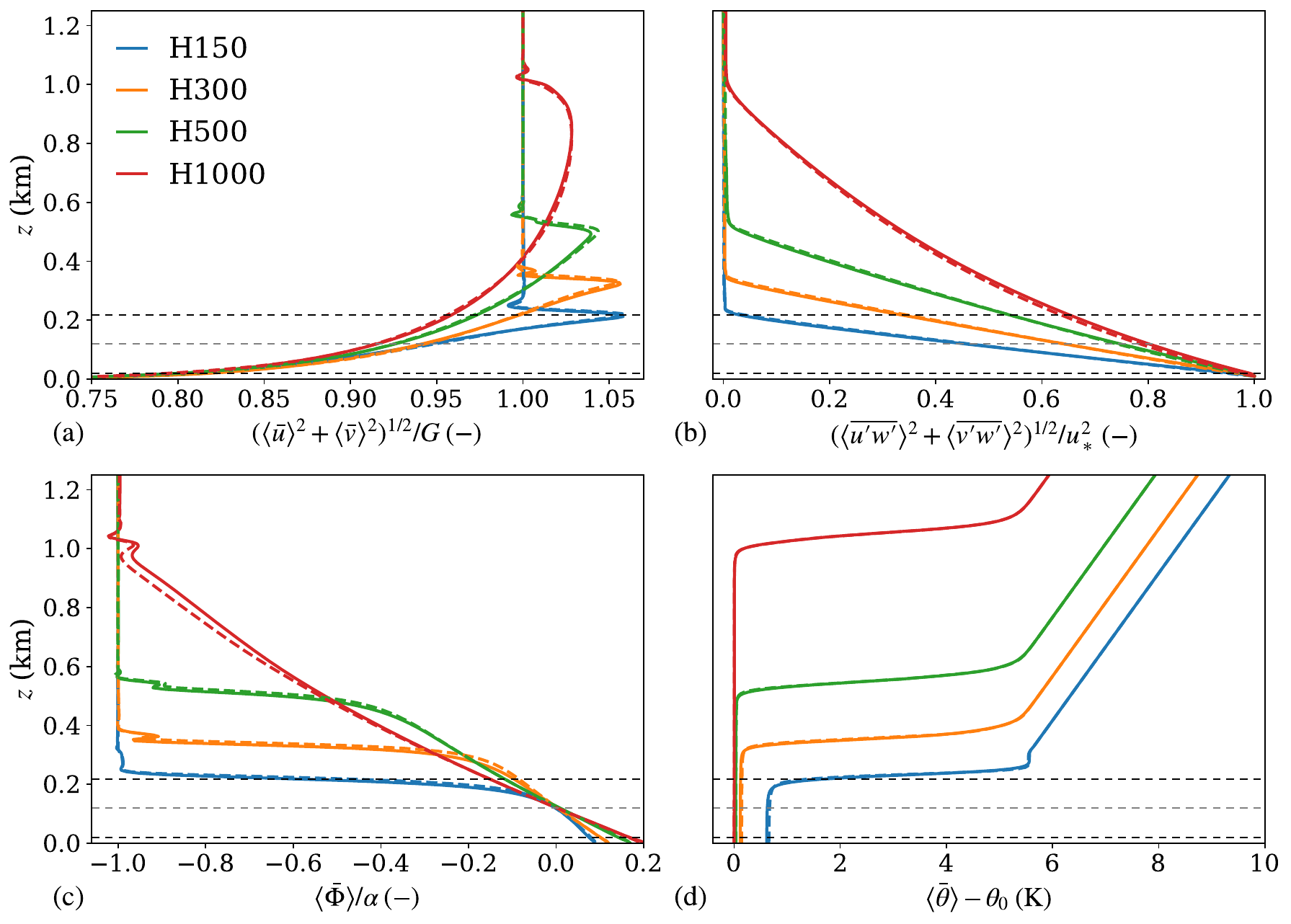}%
	\caption{Vertical profiles of (a) velocity magnitude, (b) total shear stress magnitude, (c) wind direction and (d) potential temperature averaged along the full horizontal directions and over the last $4$ h of the simulation. The continuous lines denote the profiles used in this work while the dashed lines represent the profiles obtained with a finer horizontal grid resolution with $\Delta x/2$ and $\Delta y/2$ (i.e. the profiles used by \cite{Lanzilao2024}). Finally, the grey dashed line denotes the turbine-hub height while the black dashed lines are representative of the rotor dimension. We note that the results shown here only refer to the precursor simulations.}
	\label{fig:precursor_results}
\end{figure}

Next, we add random divergence-free perturbations with an amplitude of $0.1G$ in the first $100$ m to the vertical velocity profiles. This initial state is given as input to the precursor simulation. The flow is advanced in time for $20$ h, which is sufficient to obtain a turbulent fully developed statistically steady state \citep{Pedersen2014,Allaerts2017,Lanzilao2022b,Lanzilao2024}. Figure \ref{fig:precursor_results} illustrates vertical profiles of several quantities of interest averaged over the last $4$ h of the simulations and over the full horizontal directions. Figure \ref{fig:precursor_results}(a) shows the velocity magnitude normalized with the geostrophic wind. The boundary layer extends up to the capping inversion, which limits its growth. All velocity profiles show a common feature, that is the presence of a super-geostrophic jet near the top of the ABL, which is a typical phenomenon observed in this type of atmospheric conditions \cite{Pedersen2014,Goit2015,Allaerts2015}. Such jet is more accentuated for the H150 cases, where a stronger wind shear within the ABL is attained. Next, Figure \ref{fig:precursor_results}(b) displays the sum of the modelled and resolved shear stress magnitude, which is non-zero only below the capping inversion, with a quasi-linear profile. Further, Figure~\ref{fig:precursor_results}(c) shows the flow angle. At turbine-hub height, the flow is parallel to the $x$-direction. This is achieved by using the wind-angle controller developed and tuned by \cite{Allaerts2015}, which is designed to ensure a desired orientation of the hub-height wind direction ($\Phi_d =0^\circ$ in this case). We also observe that most of the wind-direction change occurs within the inversion layer, except for case H1000. The geostrophic wind angle, which is the angle between the surface stress and the geostrophic wind velocity, is larger for shallow boundary layers, as noted by \cite{Allaerts2017}, going from $-18.55^\circ$ to $-7.65^\circ$ in cases H150 and H1000, respectively. Finally, the thermal stratification is illustrated in Figure \ref{fig:precursor_results}(d) by means of potential temperature profiles. For sake of completeness, we also show the profiles obtained on a finer grid resolution -- i.e. the one adopted by \cite{Lanzilao2024}. Figure \ref{fig:precursor_results} shows that doubling the grid cell size in the streamwise and spanwise direction has negligible effects on the precursor results, particularly in the region where the turbines are located -- see Appendix \ref{app:domain_sensitivity} for more details. The four spin-up cases together with some parameters of interest averaged over the last $4$ h of simulation are summarized in Table \ref{table:simulation_setup}. 

\begin{table}
	\begin{center}
		\def~{\hphantom{0}}
		\begin{adjustbox}{max width=\textwidth}
			\begin{tabular}{ccccccccccc}
				\textbf{Cases}  & $\boldsymbol{H}$ \textbf{(m)} & $\boldsymbol{\Delta \theta}$ \textbf{(K)} & $\boldsymbol{\Gamma}$ \textbf{(K km$^{-1}$)} & $\boldsymbol{\Delta H}$ \textbf{(m)} & $\boldsymbol{M_\mathrm{prec}}$ \textbf{(m s$^{-1}$)} & $\boldsymbol{\mathrm{TI}_\mathrm{prec}}$ \textbf{(\%)} & $\boldsymbol{u_\star}$ \textbf{(m s$^{-1}$)} & $\boldsymbol{\alpha (^\circ)}$ & \textbf{\textit{Fr} (--)} & $\boldsymbol{P_N}$ \textbf{(--)} \\[7pt]
				
				H150     & 203  & 4.72 & 4 & 52 & 9.44 & 3.09 & 0.276 & -18.55 & 1.58 & 3.43    \\
				H300     & 319  & 5.18 & 4 & 74 & 9.40 & 3.37 & 0.279 & -12.74 & 1.25 & 2.35 \\
				H500     & 507  & 5.29 & 4 & 91  & 9.25 & 3.76 & 0.276 & -9.14 & 1.00 & 1.56 \\
				H1000     & 1001  & 5.33 & 4 & 99  & 9.14 & 4.04 & 0.274 & -7.65 & 0.73 & 0.82 \\
			\end{tabular}
		\end{adjustbox}
		\caption{Overview of the spin-up cases used to drive the wind-farm simulations. The parameters are averaged over the last 4 h of the spin-up phase and include the capping-inversion height $H$, the capping-inversion strength $\Delta \theta$, the free atmosphere lapse rate $\Gamma$, the capping-inversion thickness $\Delta H$, the velocity magnitude measure at hub height $M_\mathrm{prec}$, the turbulence intensity measured at hub height $\mathrm{TI}_\mathrm{prec}$, the friction velocity $u_\star$, the geostrophic wind angle $\alpha$, the Froude number \textit{Fr} and the $P_N$ number. Note that the parameters $H$, $\Delta \theta$, $\Gamma$ and $\Delta H$ have been estimated by fitting the spin-up profiles averaged over the last 4 h of the precursor simulations with the \cite{Rampanelli2004} model.}
		\label{table:simulation_setup}
	\end{center}
\end{table}

The four spin-up cases together with some parameters of interest averaged over the last $4$ h of simulation are summarized in Table \ref{table:simulation_setup}. We remark that the capping-inversion height moves upward during the spin-up phase, particularly for the shallow boundary-layer cases. For instance, the H150 cases show a growth of $53$ m on average over the $20$~h of spin-up. Moreover, the capping-inversion strength and thickness also reduces down to $4.72$ K and $52$ m, respectively. For the H1000 cases, the vertical potential-temperature profile shows only very minor changes from its initial state. As a result of these changes, it is important to note that cases H150, H300, H500 and H1000 have an effective capping-inversion height of $203$, $319$, $507$ and $1001$ m.

\subsection{Wind-farm start-up phase}\label{sec:windfarm_startup}
The four precursor fields previously discussed are now used to drive four simulations in the main domain, where the wind farm actively imposes a drag force on the flow. However, before collecting flow statistics over time, a second spin-up phase is required. In fact, the flow has to adjust to the presence of the farm in the main domain before reaching a new statistically-steady state. \cite{Lanzilao2024} have shown that 1 hour of wind-farm spin-up time suffices for the flow to adjust to the farm drag force. However, the domain length adopted in this work is more than twice than the one used by \cite{Lanzilao2024}. Hence, we fix the duration of the wind-farm start-up phase to $5.5$ hours, which corresponds to roughly 12 and 1.7 wind-farm and domain flow-through times, respectively. Next, we switch off the wind-angle controller in the precursor domain and we collect statistics during a time window of $2$~h.

\section{Sensitivity of the wind-farm wake behaviour to the capping-inversion height}\label{sec:sensitivity_atm}
The sensitivity of the flow blockage and wind-farm performance to the thermal stratification above the ABL have been already analyzed in details by \cite{Lanzilao2024}, who adopted the same wind farm used in this study. Therefore, the focus of this section is solely on the wake properties and its recovery mechanisms as the height of the capping inversion varies. We recognize that other parameters that define the thermal stratification above the ABL may also have an influence on the wake behaviour. An example is discussed in Appendix \ref{app:ci_thickness}. Here, Section~\ref{sec:flow_physics} presents a qualitative analysis of the wake behaviour when $H$ varies. In Section~\ref{sec:wake_properties} we perform a quantitative analysis by introducing a simple fitting model for inferring quantities such as the wake strength, deflection and width along the streamwise direction. Next, a momentum budget analysis is performed in Section \ref{sec:momentum_analysis}, where we discuss in details the dominant terms that replenish the wind-farm wake. Finally, we investigate in Section \ref{sec:wake_convergence} the flow behaviour within the wake region and at its side. We remark that the results shown in the remainder of the text are time-averaged over the last $2$ hours of simulation time. Further, given a tip height of $218$~m, we will often refer to cases H150 and H300 as shallow boundary-layer cases, while H500 and H1000 will often be denoted as deep boundary-layer cases. 

\begin{figure}
	\centering
	\includegraphics[width=1.\textwidth]{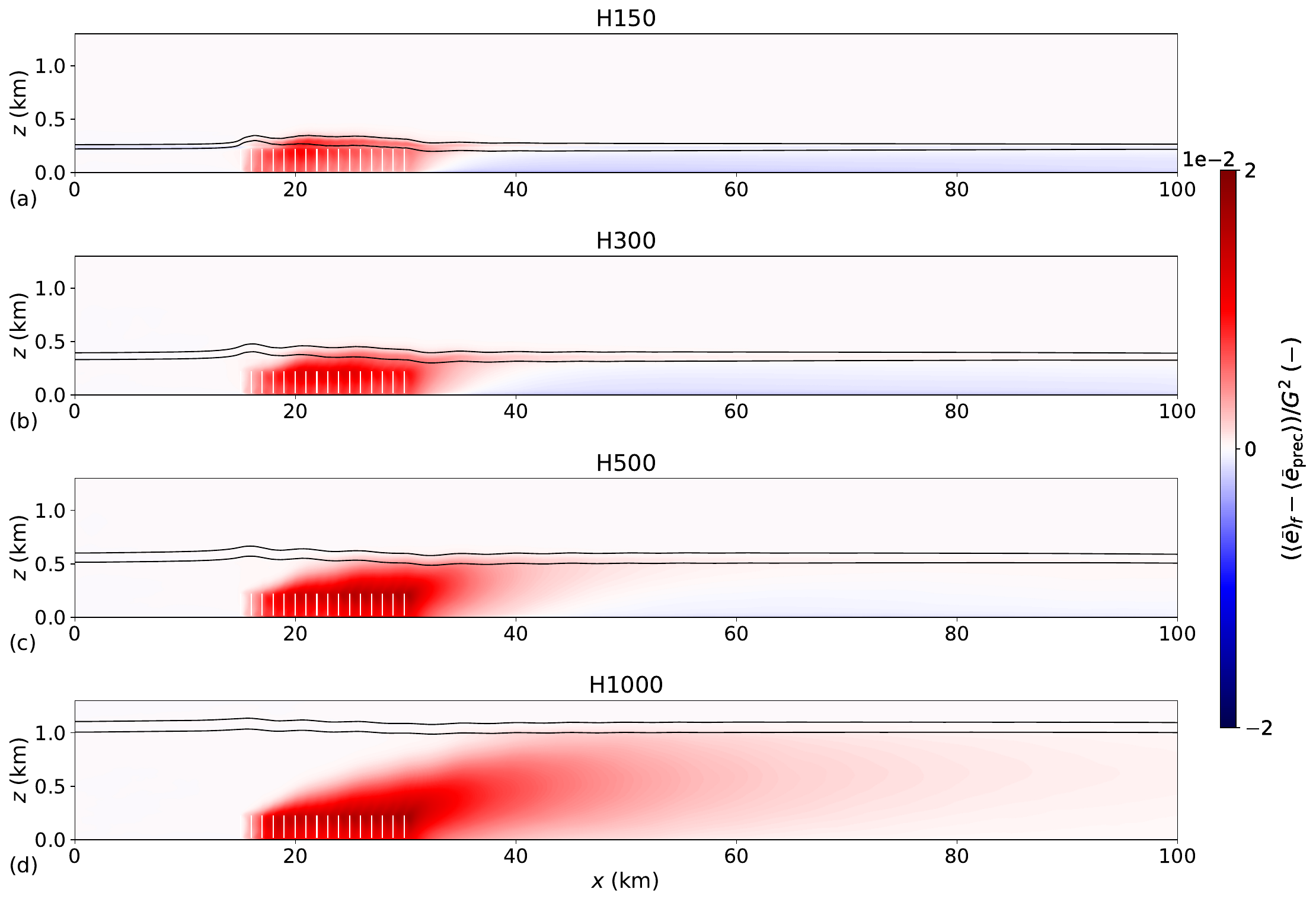}
	
	\caption{Contours of the time-averaged turbulent kinetic energy perturbation with respect to the precursor simulation taken in an $x$--$z$ plane further averaged along the farm width in the spanwise direction for cases (a) H150, (b) H300, (c) H500 and (d) H1000. The black lines represent the bottom and top of the inversion layer computed with the \cite{Rampanelli2004} model. Finally, the location of the turbine-rotor disks is indicated with vertical white lines.}
	\label{fig:xz_slices_tke}
\end{figure}

\subsection{Flow physics}\label{sec:flow_physics}
\begin{figure}
	\centering
	\includegraphics[width=1.\textwidth]{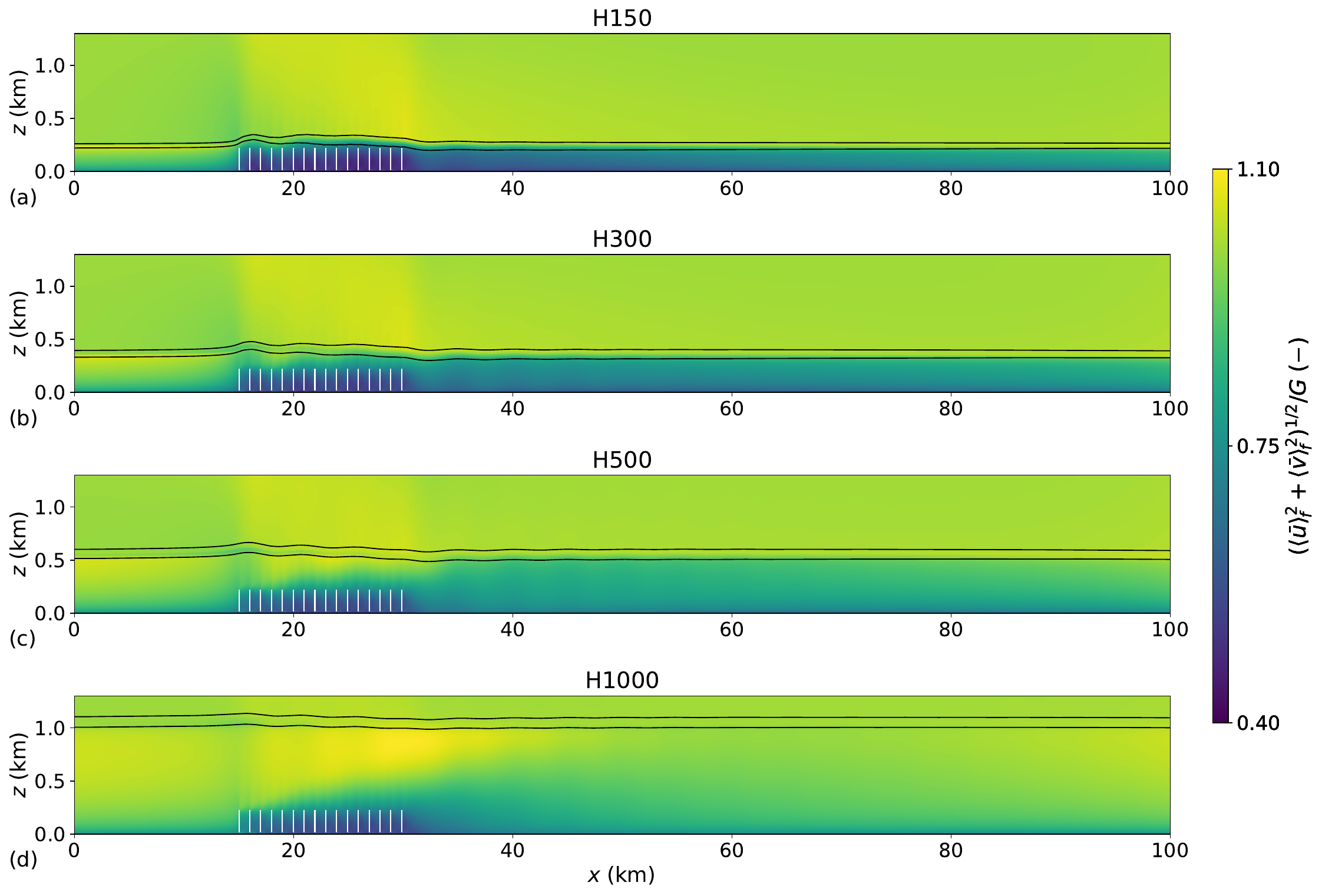}
	\caption{Contours of the time-averaged horizontal velocity magnitude in an $x$--$z$ plane further averaged along the farm width in the spanwise direction for cases (a) H150, (b) H300, (c) H500 and (d) H1000. The black lines represent the bottom and top of the inversion layer computed with the \cite{Rampanelli2004} model. Finally, the location of the turbine-rotor disks is indicated with vertical white lines.}
	\label{fig:xz_slices_m}
\end{figure}

We start our analysis with Figure \ref{fig:xz_slices_tke}, which shows a side view of turbulent kinetic energy (TKE) perturbation with respect to the precursor simulation averaged in the $y$ direction along the width of the farm, together with the base and top of the inversion layer computed by fitting the LES data with the \cite{Rampanelli2004} model. In the H150 and H300 cases, the vicinity of the capping inversion to the turbine-tip height limits the flow development in the vertical direction. Moreover, the capping inversion behaves as a pliant surface, limiting flow entrainment from the free atmosphere. Further, the high stability attained within the inversion layer dampens turbulence. Consequently, Figure~\ref{fig:xz_slices_tke}(a,b) displays relatively high values of TKE within the farm which decays downwind of the last row of turbines. Moreover, the reduced wind sheer in the wake region causes the TKE to drop below the values attained in the precursor simulations. The combination of these effects lead to very strong velocity deficit in the farm wake and a slow recovery rate, as shown in Figure~\ref{fig:xz_slices_m}(a,b). The increase in the $H/z_\mathrm{tip}$ ratio for cases H500 and H1000 allows for higher TKE values and the development of an internal boundary layer, as shown in Figure~\ref{fig:xz_slices_tke}(c,d). The latter enhances vertical turbulent transport of momentum, therefore giving rise to lower wake deficit and a faster recovery rate, as visible in Figure \ref{fig:xz_slices_m}(c,d). Moreover, case H1000 shows the formation of a high speed channel between the tip height and the capping-inversion base which further increases the TKE and flow mixing.
\cite{Stieren2022} mentioned that a wind-farm operating in the wake of an upstream one is negatively affected in terms of power output, but has a higher wake efficiency due to the increased turbulence level in the incoming inflow. Figure~\ref{fig:xz_slices_tke} shows that this is the case only in deep boundary layers and only when the farm is less distant than $10$ to $20$ km, as the TKE decays much faster than the velocity deficit. The latter, shown in Figure~\ref{fig:xz_slices_m}, illustrates that the wake is not fully replenished in all cases, meaning that it would impact on downwind farms located more than $70$~km downstream.

\begin{figure}
	\includegraphics[width=1\textwidth]{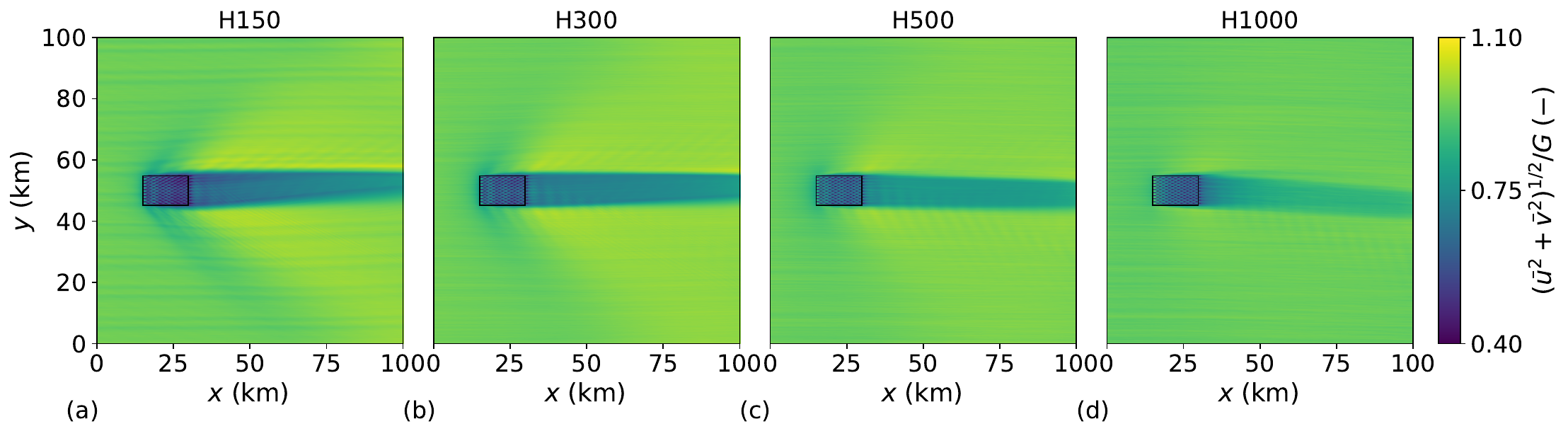}
	\includegraphics[width=1\textwidth]{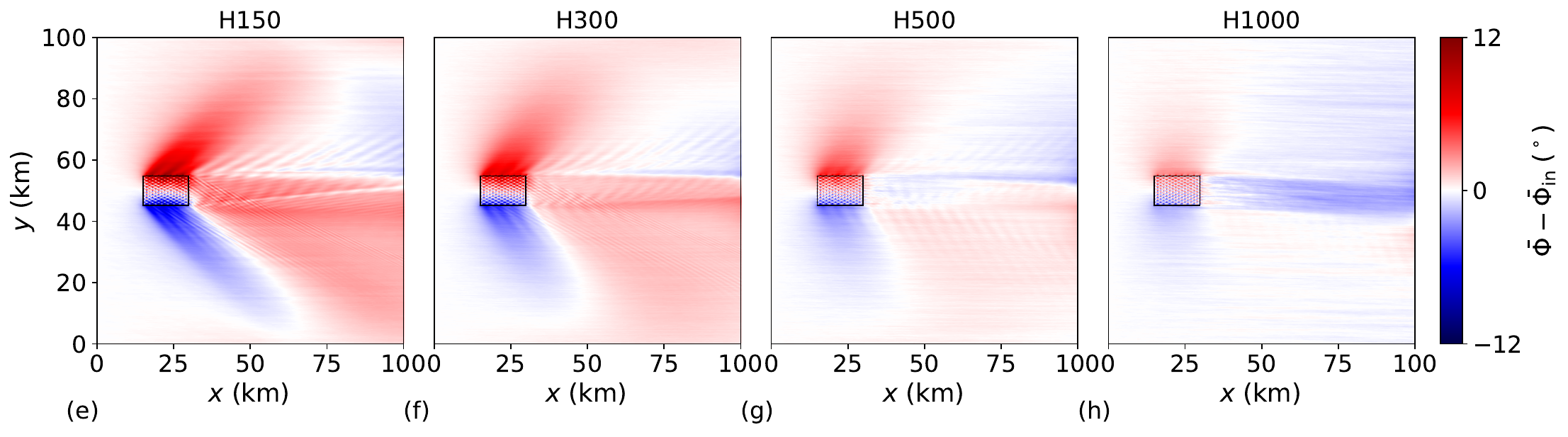}
	\includegraphics[width=1\textwidth]{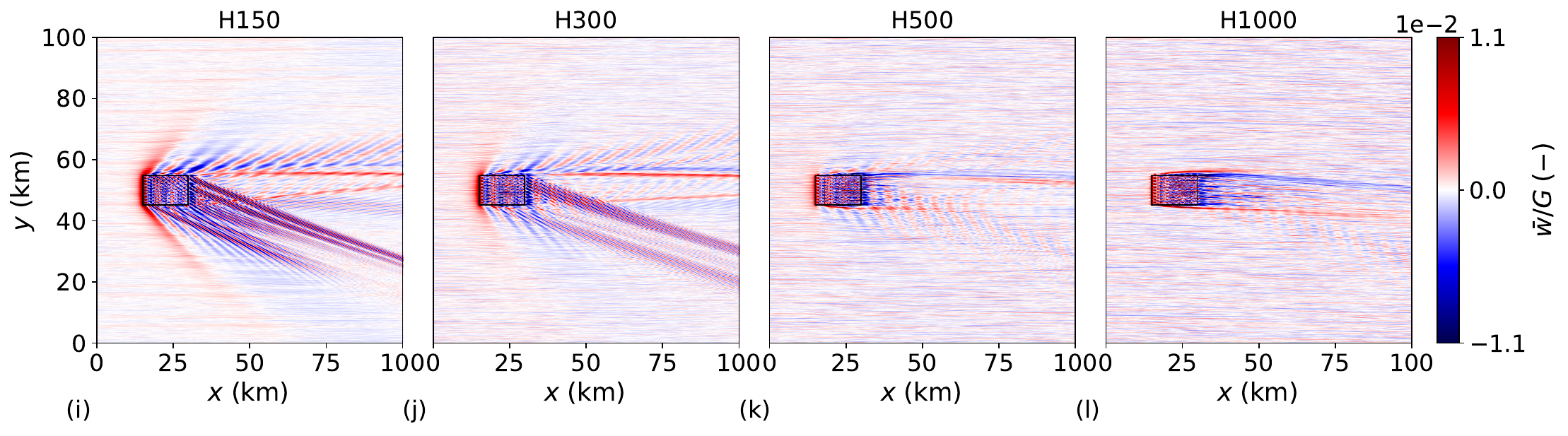}
	\includegraphics[width=1\textwidth]{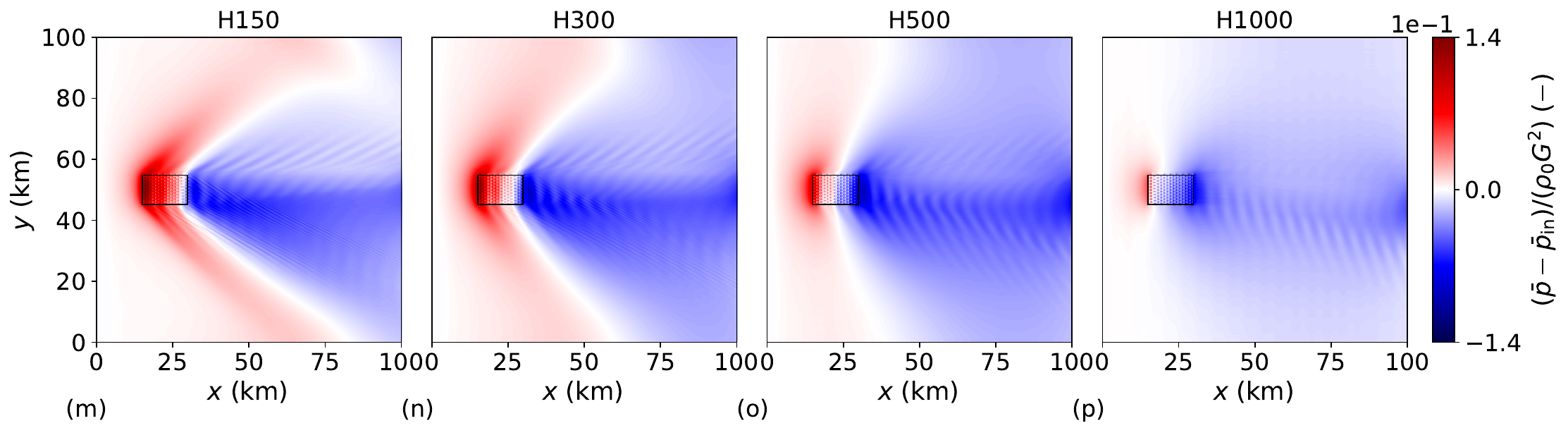}
	\includegraphics[width=1\textwidth]{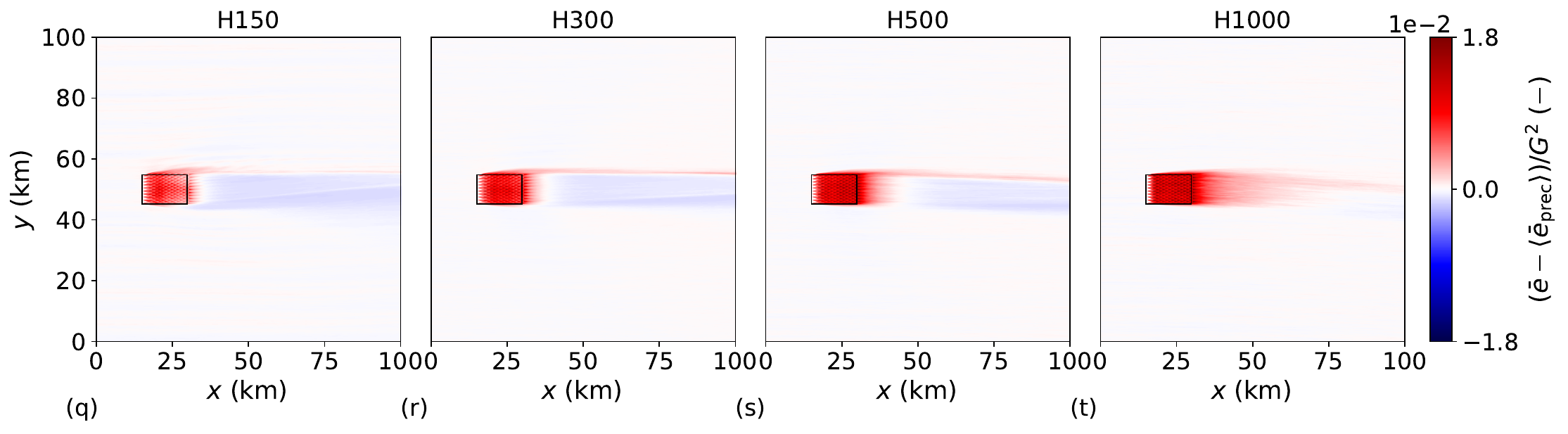}
	
	\caption{Contours of the time-averaged (a-d) horizontal velocity magnitude, (e-h) flow angle, (i-l) vertical velocity, (m-p) pressure perturbation with respect to the inflow and (q-t) turbulent kinetic energy perturbation with respect to the precursor simulation in an $x$--$y$ plane taken at hub height for cases (a,e,i,m,q) H150, (b,f,j,n,r) H300, (c,g,k,o,s) H500 and (d,h,l,p,t) H1000. The location of the wind farm is indicated with the black rectangle. Finally, we note that the last 10 km of domain in the streamwise direction are not displayed as the solution is influenced by the fringe forcing in that region.}
	\label{fig:xy_slices}
\end{figure}

Next, Figure~\ref{fig:xy_slices}(a-d) shows a top view taken at hub height of horizontal velocity magnitude. First, we observe that in all cases the velocity deficit does not seem to spread horizontally. A similar behaviour can be observed in numerical simulations and SAR images, such as Figure~\ref{fig:sar_wake} \citep{Maas2022,Baas2022,Finseraas2023}. Therefore, the assumption that the velocity deficit spreads linearly, which is typically made in analytical wake models, seems to not hold for wind-farm wakes. As shown previously, the vicinity of the capping inversion to turbine tip height in cases H150 and H300 limits the flow development in the vertical direction. Therefore, to conserve mass, high speed regions form at the sides of the wake. The same phenomenon was also observed by \cite{Lanzilao2024}. The strong flow deceleration attained in these two cases decreases the Coriolis force magnitude, which scales linearly with the wind speed. Hence, the wake turns towards the direction of the background pressure gradient, therefore undergoing an anticlockwise flow rotation. In the H500 and H1000 cases, the wake recovery rate is higher, enhanced by the vertical turbulent entrainment of momentum. The latter adds clockwise-turning flow from aloft into the wake region. Therefore, a clockwise wake deflection is observed, which is dictated by the wind veer present in the background flow. This behaviour is also illustrated in Figure~\ref{fig:xy_slices}(e-h), which displays the flow angle at hub height. These subplots clearly show that effects of Coriolis forces and wind veer are opposed. Whether one dominates over the other depends upon the capping-inversion height. Moreover, we can also conclude that, in the absence of wind veer in the background flow, the wind-farm wake would be always deflected anticlockwise in the Northern Hemisphere. We note that similar results have been observed for both single-turbine and wind-farm wake simulations \citep{Vanderlaan2017b,Howland2018,Heck2024}.

The typical signature of trapped waves is visible in the vertical velocity field shown in Figure~\ref{fig:xy_slices}(i-l), particularly in the shallow boundary layer cases where the capping inversion is closer to hub height. The V-shaped pattern attained in sub-critical flows, also visible in the pressure perturbation fields shown in Figure~\ref{fig:xy_slices}(m-p), has been already observed and analyzed in details by \cite{Allaerts2019} and \cite{Lanzilao2024}. However, similarly to \cite{Lanzilao2024}, we notice the presence of slanted waves with a high wave-number which cross the full domain in the streamwise direction, particularly visible in cases H150 and H300. Finally, we note in Figure~\ref{fig:xy_slices}(i,j) the presence of a region of positive vertical velocity on the left edge of the wake. This denotes a region of flow convergence, which will be discussed in Section~\ref{sec:wake_convergence}. To conclude, Figure~\ref{fig:xy_slices}(q-t) illustrates the TKE perturbation with respect to the precursor simulation at hub height. As expected, a higher TKE level is obtained in deeper boundary layers. Moreover, the TKE in the wake of the farm is smaller than the background value, as also shown in Figure~\ref{fig:xz_slices_tke}(a,b). This is a consequence of the reduced velocity shear in the farm wake, which therefore diminishes the production of turbulence. Similar results have been observed in wind tunnel experiments and mesoscale models \citep{Chamorro2009,Fitch2012}. Finally, similar to \cite{Maas2023}, we notice a higher level of TKE on the left side of the wake, i.e. at $y=55$ km. 

\begin{figure}
	\centering
	\includegraphics[width=0.7\textwidth]{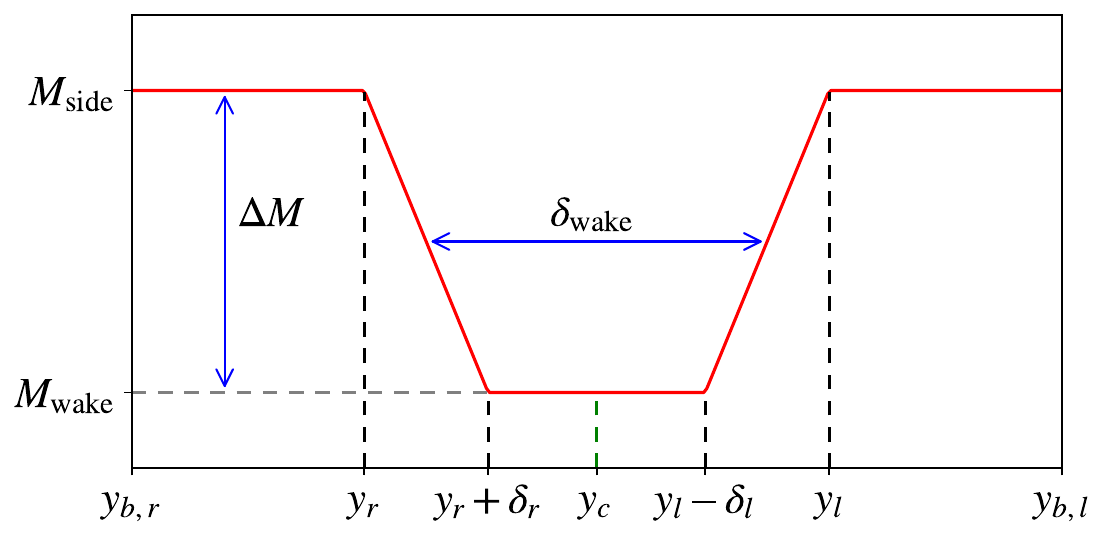}
	\caption{Sketch of the fitting function $F(y)$ and its parameters -- see Equation \ref{eq:fitting_model}.}
	\label{fig:fit_model_sketch}
\end{figure}

\begin{figure}
	\includegraphics[width=1\textwidth]{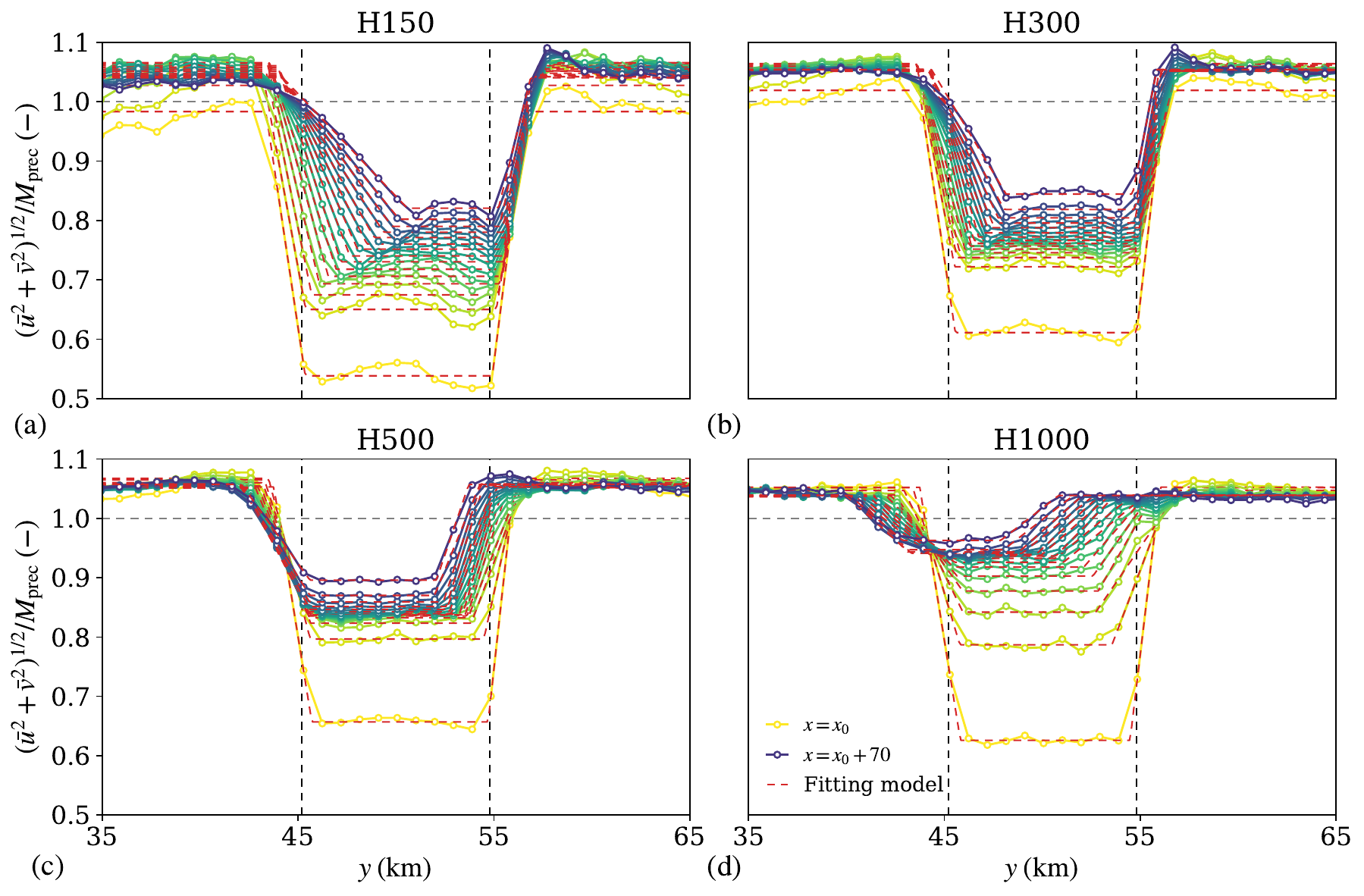} 
	
	\caption{Time-averaged horizontal velocity magnitude as a function of the spanwise direction obtained at $x=x_0$ (brightest yellow) and $x=x_0+70$ (darkest blue), with increments of $5$ km in between, for cases (a) H150 and (b) H1000. The red dashed lines illustrates the results of our fitting model for each streamwise location. Finally, the vertical dashed black lines denote the location of the first- and last-column turbines. We remark that we define the regions around $y=45$ km and $y=55$ km as the right and left edges, respectively. Moreover, $x_0=30$ km denotes the beginning of the wake region.}
	\label{fig:fitting_model}
\end{figure}

\subsection{Wake properties}\label{sec:wake_properties}
We have seen above that the wake region is heavily dependent upon the capping-inversion height. Here, we will rigorously define the wake edges and center, which we use to evaluate the wake width, strength and deflection. To do so, we propose a simple function which we use to fit the velocity magnitude profiles along the spanwise direction, which reads as follows
\begin{equation}
	F(y) = 
	\begin{cases}
		M_\mathrm{side},& \text{if } y_{b,r} \leq y < y_r\\
		M_\mathrm{side} - \Delta M (y-y_r)/\delta_r ,& \text{if } y_r \leq y < y_r + \delta_r \\
		M_\mathrm{wake} ,& \text{if } y_r + \delta_r \leq y < y_l - \delta_l \\
		M_\mathrm{wake} + \Delta M (y-y_l+\delta_l)/\delta_l,& \text{if } y_l - \delta_l \leq y < y_l \\ 
		M_\mathrm{side},& \text{if } y_l \leq y < y_{b,l}
	\end{cases}
	\label{eq:fitting_model}
\end{equation}
where $M_\mathrm{side}$ and $M_\mathrm{wake}$ denote the velocity magnitude attained at the sides and in the wake of the farm while $y_r$ and $y_l$ denote the right and left wake-edge locations, located in the region around $y=45$ and $y=55$ km, respectively. We remark that due to the convention used, the right side of the wake corresponds to the left side of the figure, and vice versa. Moreover, we note that the velocity $M_\mathrm{side}$ is measured in a region that extends $10$~km sideways of the farm location, so that its value does not depend upon the spanwise domain dimension. Hence, $y_{b,r}$ and $y_{b,l}$ assume the value of $35$ and $65$ km, respectively. Further, $\delta_r$ and $\delta_l$ represent the distance over which the velocity decays from $M_\mathrm{side}$ to $M_\mathrm{wake}$ while $\Delta M = M_\mathrm{side} - M_\mathrm{wake}$ indicates the velocity deficit with respect to the wind speed at the wake sides. Following \cite{Pope2000}, we define the farm wake width as $\delta_\mathrm{wake} = y_l - \delta_l/2 - (y_r + \delta_r/2)$. Further, we define the wake center as the middle point of the region where $M_\mathrm{wake}$ is attained, which is formally defined as $y_c=(y_l-\delta_l+y_r+\delta_r)/2$. The fitting function $F(y)$ reported in Equation \ref{eq:fitting_model} together with a visual representation of its parameters is shown in Figure \ref{fig:fit_model_sketch}.

The model reported in Equation \ref{eq:fitting_model} is adequate to fit the velocity magnitude profiles along the spanwise direction downwind of a wind farm. These profiles are shown in Figure \ref{fig:fitting_model} at various locations downwind of the farm (i.e. from $x=x_0$ to $x=x_0+70$ with increments of $5$~km). Figure \ref{fig:fitting_model}(a,b) shows the profiles for cases H150 and H300. Here, we can observe strong velocity deficits within the first $10$~km, where the profiles assume a top-hat shape with peaks in the location of the first and last column of turbines. This behaviour may be caused by the strong flow redirection at the farm sides, which yaws the first and last column of turbines up to 8$^\circ$ in case H150, allowing for a higher energy extraction and therefore causing higher velocity deficit at $y=45$~and $y=55$~km \citep{Lanzilao2024}. In these cases, the wake turns anticlockwise. Therefore, the velocity gradient on the right edge of the wake (i.e at $y=45$~km) diminishes along the streamwise direction, while it remains very strong on the left edge (i.e at $y=55$~km). This phenomenon happens when the wake turns in the opposite direction of the wind veer of the background flow and will be further discussed in Section \ref{sec:wake_convergence}. Next, Figure \ref{fig:fitting_model}(c,d) shows the results for the deep boundary layer cases, i.e. H500 and H1000. The velocity profiles fully resemble a top-hat shape with smooth edges. The wake turns clockwise in these cases and the velocity speed-up at the wake sides is less pronounced. The velocity magnitude profiles discussed above are then fitted using the model reported in Equation~\ref{eq:fitting_model}. The fitting is performed using the \textit{curve\textunderscore fit} method provided by the Scipy library, adopting the trust region reflective algorithm \citep{Scipy2020}. The results are shown in Figure~\ref{fig:fitting_model}, where the red dashed lines illustrate the fitted velocity profiles. Here, we can observe that the model captures the wake location and strength very well.

\begin{figure}
	\includegraphics[width=0.5\textwidth]{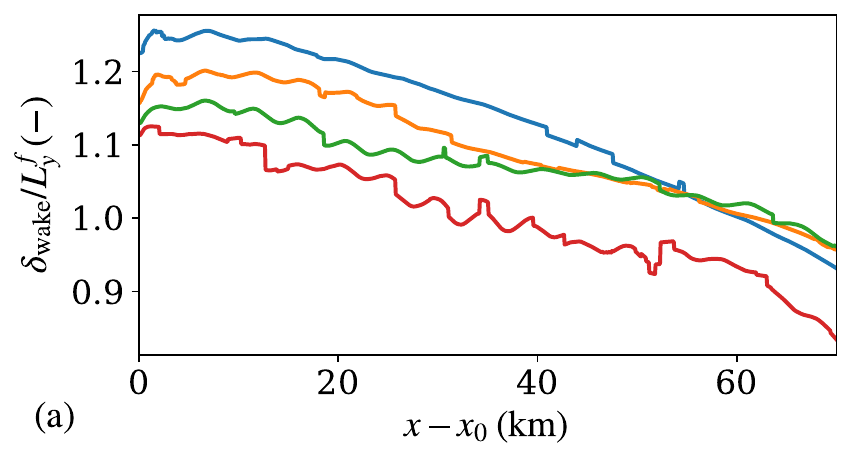} 
	\includegraphics[width=0.5\textwidth]{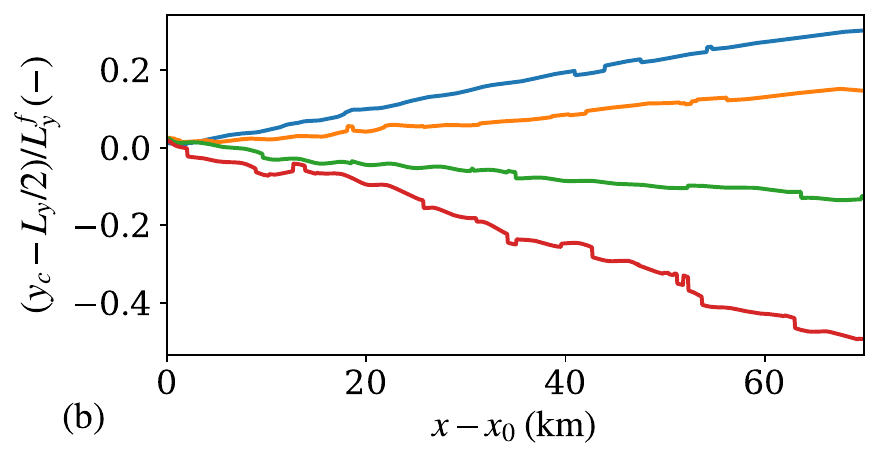} 
	\includegraphics[width=0.5\textwidth]{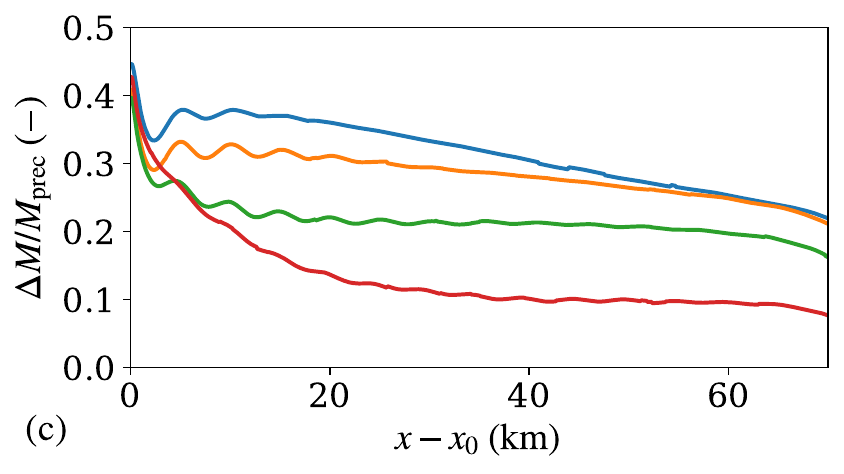} 
	\includegraphics[width=0.5\textwidth]{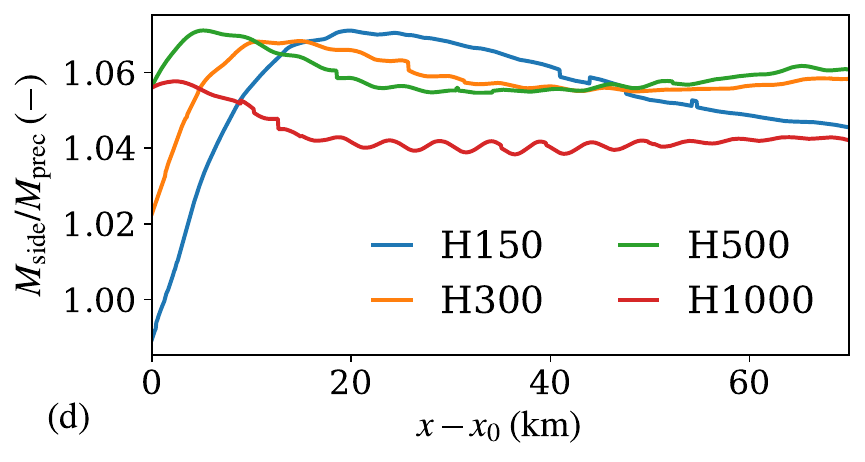} 
	
	\caption{(a,b) Wake width and wake center normalized with the farm width, (c,d) velocity deficit with respect to $M_\mathrm{side}$ and velocity magnitude at the sides of the wake normalized with the velocity magnitude obtained in the precursor domain at hub height. All quantities are predicted by the fitting model shown in Equation~\ref{eq:fitting_model}. Moreover, the $x$-axis is rescaled with $x_0=30$ km which denotes the beginning of the wake region.}
	\label{fig:fit_wake_properties}
\end{figure}

The wake width and center, the velocity deficit and the velocity at the wake sides all evaluated at hub height and estimated using the fitting model reported in Equation \ref{eq:fitting_model}, are shown in Figure \ref{fig:fit_wake_properties}. The wake width is shown in Figure \ref{fig:fit_wake_properties}(a). Here, we can observe a width $10$ to $20$\% higher than $L_y^f$ at the beginning of the wake region. Interestingly, the wake narrows in all cases along the downstream direction. At $x-x_0=65$ km, cases H150, H300 and H500 show a wake-to-farm width ratio of roughly one while case H1000 displays a narrower wake. The wake narrowing along the streamwise direction can also be observed in SAR images and in the results shown by \cite{Baas2022,Baas2023,Maas2022} and \cite{Maas2023}. The wake center is displayed in Figure \ref{fig:fit_wake_properties}(b). Here, we observe how the capping-inversion height influences the wake deflection, generating a clear distinction between shallow and deep boundary layers. The strongest wake deflection is observed in case H1000, where the wake center deviates of about $45$\% the wake width $65$ km downstream of the farm. 

\begin{figure}
	\includegraphics[width=0.5\textwidth]{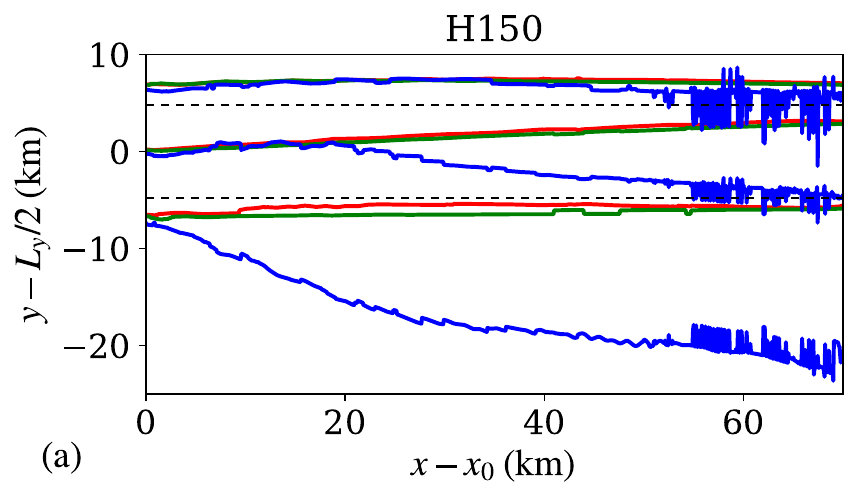} 
	\includegraphics[width=0.5\textwidth]{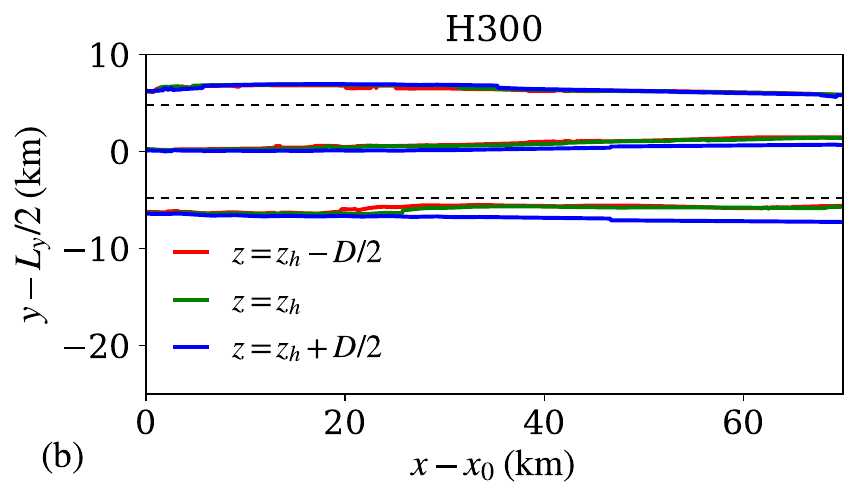} 
	\includegraphics[width=0.5\textwidth]{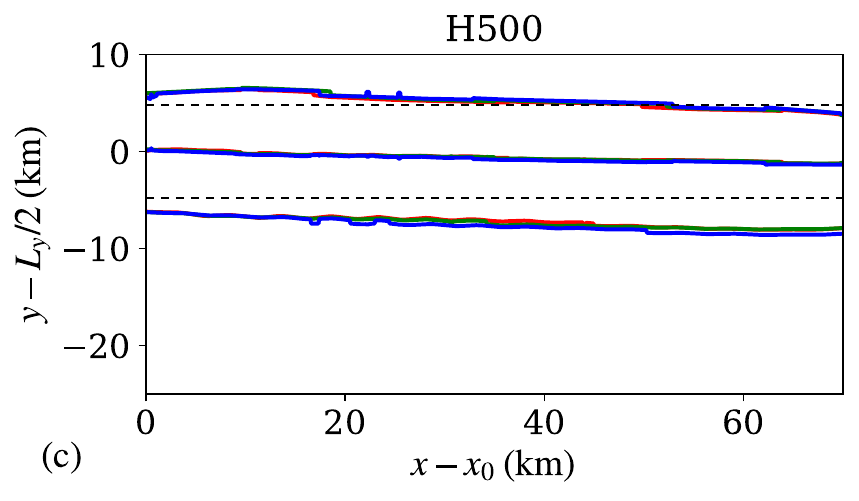} 
	\includegraphics[width=0.5\textwidth]{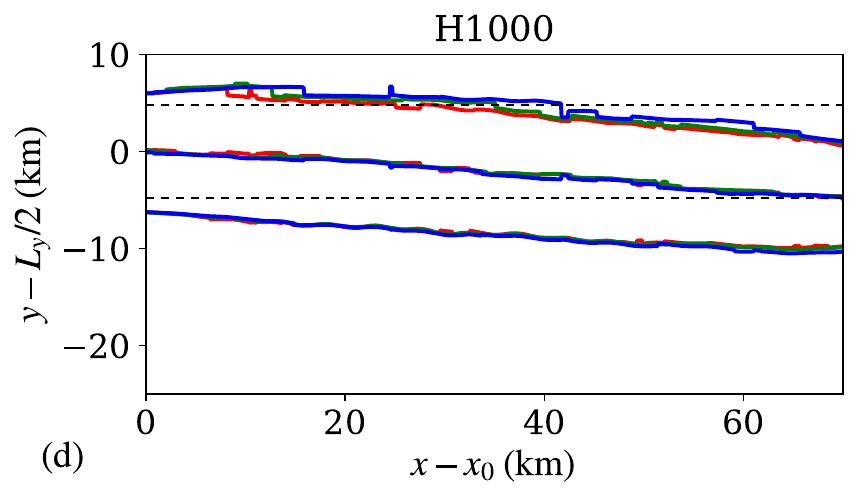} 
	
	\caption{(a) Wake edges and center as a function of height for cases (a) H150, (b) H300, (c) H500 and (d) H1000 predicted by the fitting model shown in Equation~\ref{eq:fitting_model}. The black horizontal dashed lines denote the location of the first and last turbine column. Moreover, the $x$-axis is rescaled with $x_0=30$ km which denotes the beginning of the wake region.}
	\label{fig:fit_wake_height}
\end{figure}

Next, Figure~\ref{fig:fit_wake_properties}(c) illustrates the velocity deficit with respect to the velocity measured at the sides of the wake and further normalized with the precursor velocity at hub height. In all cases, the wake shows a very fast recovery rate within the first $5$ km. Afterwards, the velocity deficit attains a very slow recovery, except for case H1000 where $\Delta M$ keeps decreasing up to $40$~km downwind of the farm. We note that the ratio $\Delta M/M_\mathrm{prec}$ measures approximately $0.25$ at the end of the wake region in cases H150 and H300, confirming that the wake is not replenished. Moreover, we remark the presence of oscillations in the $\Delta M$ profiles with decreasing amplitude for increasing values of $H$. These oscillations are related to the flow divergence--convergence generated by interfacial waves (see Figure \ref{fig:xy_slices}(i-l)), and therefore are more accentuated in shallow boundary layers. Finally, Figure~\ref{fig:fit_wake_properties}(d) shows the velocity at the sides of the wake normalized with $M_\mathrm{prec}$. \cite{Lanzilao2024} have shown that shallow boundary layers cause the flow to speed-up at the farm sides. The same phenomenon is visible here. For instance, in case H150, $M_\mathrm{side}/M_\mathrm{prec}$ goes from $0.99$ to $1.07$ in the first $20$ km downwind of the farm. A high capping inversion allows the internal boundary layer to grow vertically, resulting in lower $M_\mathrm{side}/M_\mathrm{prec}$ ratios. Finally, we note that the maximum $M_\mathrm{side}/M_\mathrm{prec}$ value shifts downstream as $H$ decreases. 

Up to this point, we have only investigated the wake properties at hub height. Looking at the wake evolution at different heights, Figure~\ref{fig:fit_wake_height} shows the wake edges $y_r+\delta_r/2$ and $y_l-\delta_l/2$ together with the wake center at the height of $z_h-D/2$, $z_h$ and $z_h+D/2$, for all cases. While the wake turns anticlockwise at hub height for case H150, we can observe a strong clockwise rotation of the velocity deficit at tip height. This effect is caused by the strong wind veer attained within the capping inversion, which is very near to tip height in this case. Moreover, the noise observed in the blue lines around $x=90$ km is due to the vicinity of the capping inversion, where the velocity deficit assumes a different profile than the one our model can fit. In all other cases, the wake edges and center do not differ substantially across the rotor height. Moreover, we also note that the wake recovery rate does not change across the rotor height (not shown).

\begin{figure}
	\includegraphics[width=1\textwidth]{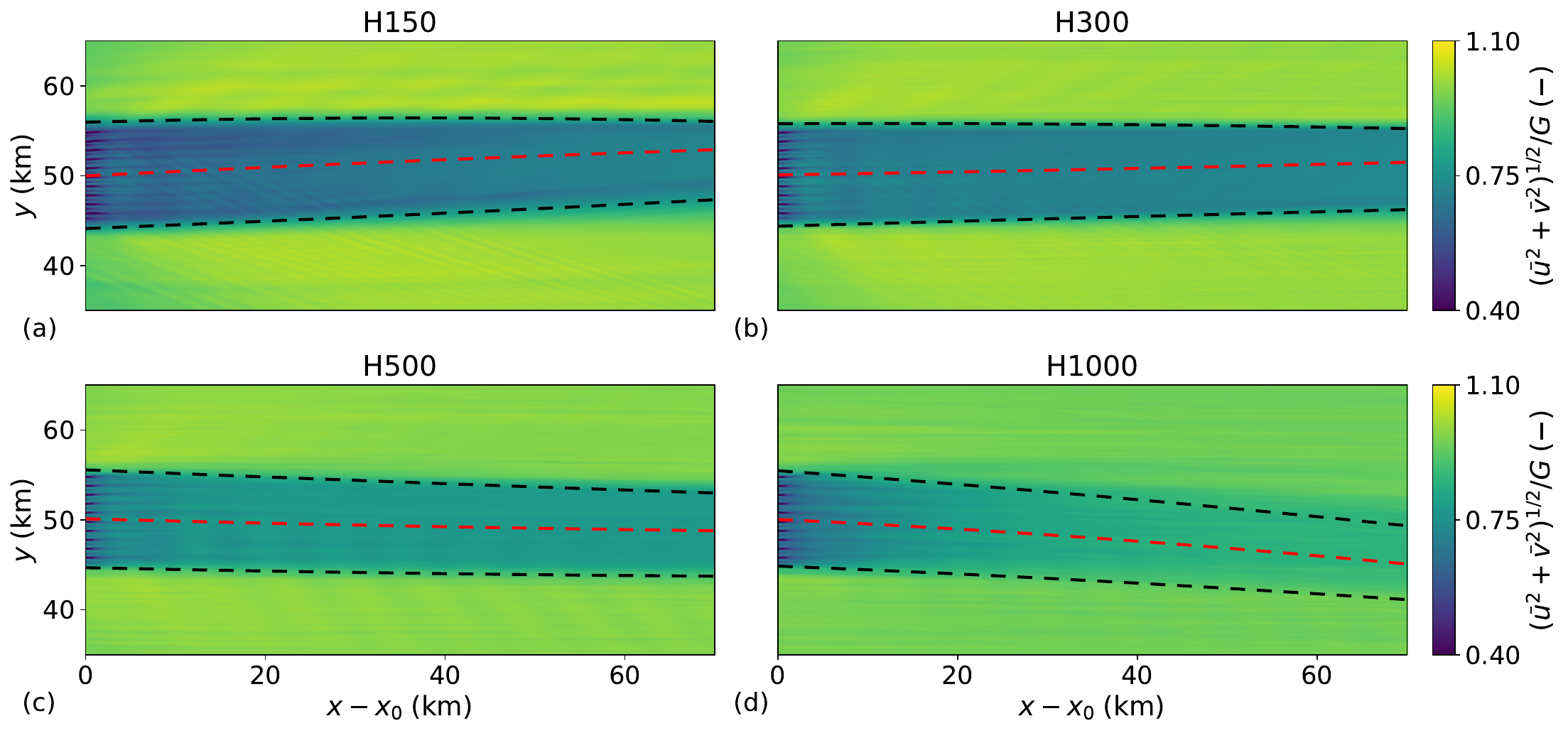}  
	
	\caption{Contours of the time-averaged horizontal velocity magnitude downwind of the farm for cases (a) H150, (b) H300, (c) H500 and (d) H1000 in an $x$--$y$ plane taken at hub height. The black and red dashed lines denote the wake edges and the wake center, respectively, predicted by the fitting model shown in Equation \ref{eq:fitting_model}. Moreover, the $x$-axis is rescaled with $x_0=30$ km which denotes the beginning of the wake region.}
	\label{fig:fit_flow_field}
\end{figure}

Next, we apply a quadratic fit to the profiles at hub height shown in Figure~\ref{fig:fit_wake_height}, and we superimpose them to the hub-height horizontal velocity magnitude field. The results for all cases are shown in Figure \ref{fig:fit_flow_field}. Here, we can observe that we correctly track the wake evolution along the streamwise direction. This makes the fitting model reported in Equation \ref{eq:fitting_model} a useful tool for inferring wake properties whenever a horizontal plane of velocity in the wake of a farm is available, which could be provided by either numerical simulations, lidar measurements or SAR images.

\begin{figure}
	\centering
	\includegraphics[width=0.55\textwidth]{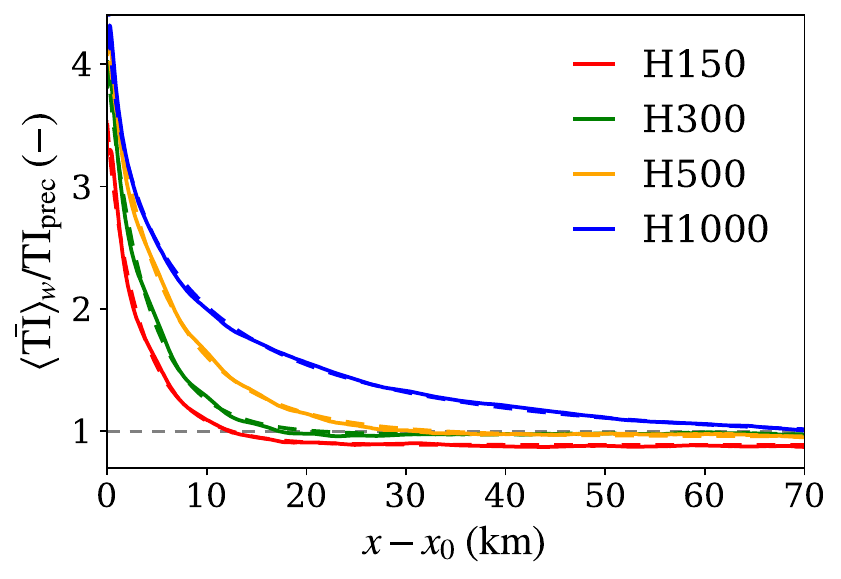}
	\caption{Time-averaged turbulence intensity further averaged over the wake width as a function of the streamwise direction further normalized with the turbulent intensity value attained in precursor simulation -- see Table \ref{table:simulation_setup}. The dashed lines in panel (b) are obtained using the expression $a_t \text{exp}(-b_t(x-x_0)^{c_t}) +d_t$, with $x_0=30$ km and $\Phi_\mathrm{TI} = [a_t,b_t,c_t,d_t]$ the set of fitting parameters, with $\Phi_\mathrm{TI}^{H150}=[8.79,0.34,0.88,2.94]$, $\Phi_\mathrm{TI}^{H300}=[10.44,0.29,0.92,3.30]$, $\Phi_\mathrm{TI}^{H500}=[12.72,0.25,0.81,3.61]$ and $\Phi_\mathrm{TI}^{H1000}=[16.21,0.39,0.52,3.59]$. We note that the $x$-axis is rescaled with $x_0=30$ km which denotes the beginning of the wake region.}
	\label{fig:x_profiles_m_ti}
\end{figure}

To conclude, we show in Figure~\ref{fig:x_profiles_m_ti} the turbulence intensity measured at hub height and as a function of the streamwise direction. We note that these profiles are further averaged along the $y$ direction over the wake width evaluated with the model reported in Equation \ref{eq:fitting_model}. For case H150, the turbulence intensity attains a constant value $20$~km downwind of the farm, while $50$~km are necessary in case H1000 for the farm-added turbulence intensity to decay. While the turbulence intensity tends to the precursor value in cases H500 and H1000, a lower value is attained in shallow boundary layer cases due to the reduced wind shear within the turbine rotor disk with respect to the background flow. Similarly to \cite{Maas2022b}, we observe that the decay of the turbulence intensity can be well approximate by the expression $a_{t} \text{exp}(-b_{t} (x-x_0)^{c_t} ) +d_{t} $, where $\Phi_\mathrm{TI} = [a_{t} ,b_{t} ,c_{t} ,d_{t}]$ represents a set of fitting parameter and $x_0=30$~km denotes the beginning of the wake region. The dashed lines in Figure \ref{fig:x_profiles_m_ti}(b) represent the fitting function. The fitting parameters for each capping-inversion height are reported in the caption of Figure \ref{fig:x_profiles_m_ti}. 

\subsection{Streamwise momentum budget analysis}\label{sec:momentum_analysis}
To highlight the dominant recovery mechanisms for the different cases, we perform a streamwise momentum budget analysis. A mass balance analysis is also performed, with results reported in Appendix \ref{app:mass_balance}. We remark that in both analyses we solely focus on the region downwind of the farm. We refer to \cite{LanzilaoPhD} and \cite{Lanzilao2024} for an in-depth analysis of the momentum and energy budget upstream and within the farm. 

The momentum analysis adopts a control volume $\Omega$ that measures $s_x D/2$ along the streamwise direction. The vertical dimension of the control volume coincides with the turbine rotor height, that is from $z_1 = z_h - D/2$ to $z_2 = z_h + D/2$. Since we focus on the momentum balance within the farm wake, the lateral extension of $\Omega$ is dictated by the wake width provided by the fitting model discussed in Section \ref{sec:wake_properties}. Hence, the lateral faces of the control volume change location along the streamwise direction and are defined as $y_1 = y_r + \delta_r/2$ and $y_2=y_l-\delta_l/2$ (i.e. the wake edges). We note that we assume the lateral edges of the control volume to be constant with height, meaning that $y_1$ and $y_2$ are not function of $z$. Figure \ref{fig:fit_wake_height} shows that this is a valid assumption, apart from case H150 where the farm wake abruptly deviates clockwise at the turbine-tip height. In summary, a generic control volume $\Omega_i$ has dimension $s_x D/2 \times \delta_\mathrm{wake}(x_i) \times D$. We remark that we denote with $x_1$, $x_2$, $y_1$, $y_2$, $z_1$ and $z_2$ the boundaries of the control volume while the $y$-$z$, $x$-$z$ and $x$-$y$ faces are denoted with $\Gamma_x$, $\Gamma_y$ and $\Gamma_z$. 

\begin{figure}
	\includegraphics[width=1\textwidth]{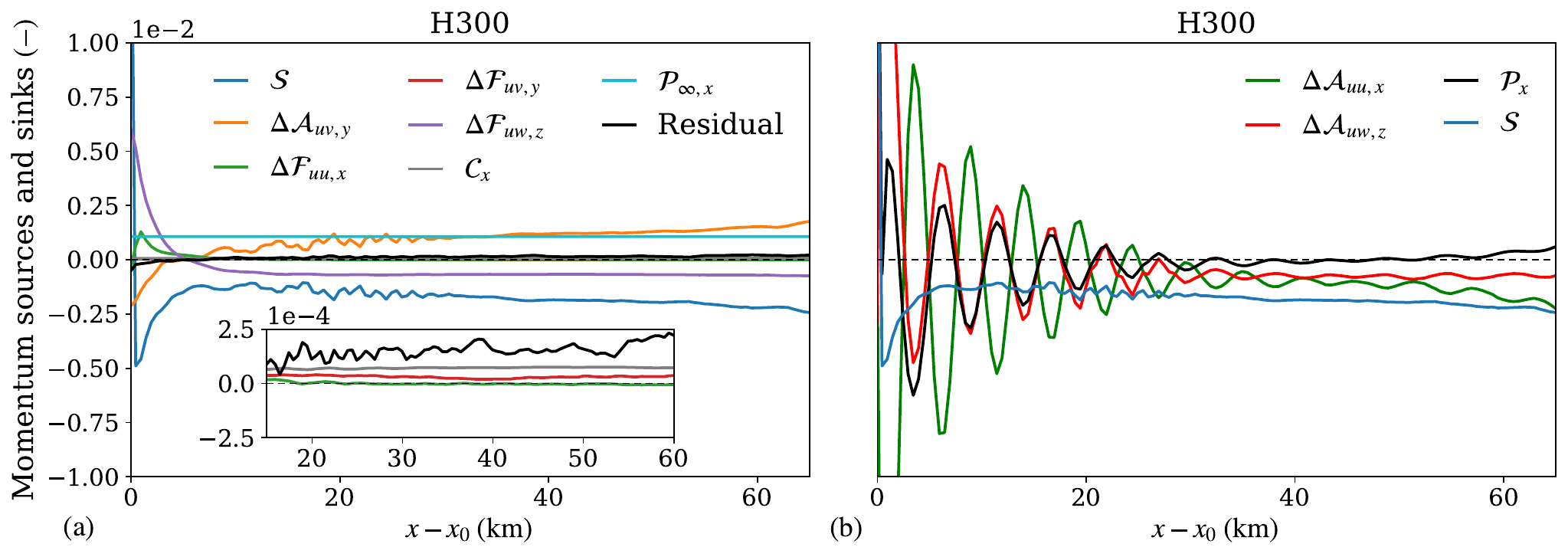} 
	\caption{Streamwise variation of momentum sources and sinks for case H300 normalized by the time averaged total wind-farm thrust force and the control volume width. Panel (a) shows all terms of Equation \ref{eq:x_mom} including $\mathcal{S}=\Delta \mathcal{A}_{uu,x} + \Delta \mathcal{A}_{uw,z} + \mathcal{P}_x$ while panel (b) shows the contribution of each added in the $\mathcal{S}$ term. We note that only the region downwind of the farm is considered, so that the term $\mathcal{F}_{t,x}$ has no contribution. Moreover, the $x$-axis is rescaled with $x_0=30$ km which denotes the beginning of the wake region.}
	\label{fig:x_mom_all_H300}
\end{figure}

The budget equation is obtained by taking a time average of the streamwise momentum equation, further integrating it over the control volume $\Omega$ to average out local oscillations. Additionally, we apply the divergence theorem, enabling us to eliminate the divergence operator and transition from a volume to a surface integral. As a result, the streamwise momentum equation for any control volume $\Omega$ reads as
\begin{equation}
	\small
	\begin{aligned}
		&\underbrace{-\biggl[ \int_{\Gamma_x} \bar{u}\bar{u} d\Gamma_x\biggl]_{x_1}^{x_2}}_{\Delta\mathcal{A}_{uu,x}} 
		\underbrace{-\biggl[ \int_{\Gamma_y} \bar{u}\bar{v} d\Gamma_y\biggl]_{y_1}^{y_2}}_{\Delta\mathcal{A}_{uv,y}} 
		\underbrace{-\biggl[ \int_{\Gamma_z} \bar{u}\bar{w} d\Gamma_z\biggl]_{z_1}^{z_2}}_{\Delta\mathcal{A}_{uw,z}} +\\[5pt]
		&\underbrace{-\biggl[ \int_{\Gamma_x} \bigl(\overline{u'u'}^r+\overline{u'u'}^\mathrm{sgs} \bigl) d \Gamma_x \biggl]_{x_1}^{x_2}}_{\Delta \mathcal{F}_{uu,x}} 
		\underbrace{-\biggl[ \int_{\Gamma_y} \bigl(\overline{u'v'}^r+\overline{u'v'}^\mathrm{sgs} \bigl) d \Gamma_y \biggl]_{y_1}^{y_2}}_{\Delta \mathcal{F}_{uv,y}}  
		\underbrace{-\biggl[ \int_{\Gamma_z} \bigl(\overline{u'w'}^r+\overline{u'w'}^\mathrm{sgs} \bigl) d \Gamma_z \biggl]_{z_1}^{z_2}}_{\Delta \mathcal{F}_{uw,z}} +\\[5pt]
		&\underbrace{\int_\Omega f_c \overline{v} d\Omega}_{\mathcal{C}_x}  \underbrace{-\int_\Omega \frac{1}{\rho_o} \frac{\partial \overline{p}^\ast}{\partial x} d\Omega}_{\mathcal{P}_x^\ast} 
		\underbrace{-\int_\Omega \frac{1}{\rho_o} \frac{\partial \overline{p}_{\infty}}{\partial x} d\Omega}_{\mathcal{P}_{\infty,x}} +
		\underbrace{\int_\Omega \overline{f}_x d\Omega}_{\mathcal{F}_{t,x}}= 0
	\end{aligned}
	\label{eq:x_mom}
\end{equation}
where $u'=u-\bar{u}$. The terms $\Delta\mathcal{A}_{uu,x}$, $\Delta\mathcal{A}_{uv,y}$ and $\Delta\mathcal{A}_{uw,z}$ represent the advection of streamwise, spanwise and vertical momentum by the streamwise velocity. The terms $\Delta \mathcal{F}_{uu,x}$, $\Delta \mathcal{F}_{uv,y}$ and $\Delta \mathcal{F}_{uw,x}$ denote the total (i.e. resolved + modelled) turbulent transport of streamwise momentum along the streamwise, spanwise and vertical directions. Further, $\mathcal{C}_x$ denotes the contribution of the Coriolis force, $\mathcal{P}_x^\ast$ and $\mathcal{P}_{\infty,x}$ represent the contribution of the pressure gradients induced by the wind-farm forcing and the background pressure gradient while $\mathcal{F}_{t,x}$ indicates the turbine forcing term. The latter term is zero out of the wind-farm region and therefore is not included in the analysis below. We note that, due to the sign convention chosen, the first six terms in Equation~\ref{eq:x_mom} are positive when the mean flow or turbulence transports more momentum in than out of the control volume and negative when the opposite occurs. For example, the term $\Delta\mathcal{A}_{uu,x}$ is positive when $\bar{u} \bar{u}$ integrated over $\Gamma_{x_1}$ is higher than when integrated over $\Gamma_{x_2}$, meaning that an increase in streamwise velocity along the $x$ direction causes this term to be negative. Finally, we note that the spanwise momentum equation will not be considered in this study since its terms are of a much smaller magnitude.

Figure \ref{fig:x_mom_all_H300}(a) shows the streamwise evolution of all terms of Equation \ref{eq:x_mom} for case H300. The close vicinity of the capping inversion to turbine tip-height dampens the vertical turbulent entrainment of momentum. In fact, $\Delta \mathcal{F}_{uw,z}$ drops abruptly in the first $5$ km, attaining a similar value to the one observed in the precursor simulation $10$~km into the wake region. This term contributes to the replenishment of the wake only in the near-wake region, where a fast recovery rate is observed -- see, e.g. Figure \ref{fig:fit_wake_properties}(c). Further, $\Delta \mathcal{F}_{uu,x}$ also attains a positive value in the near-wake region. However, this term is about three times smaller than $\Delta \mathcal{F}_{uw,z}$ and it becomes negligible $5$ km downstream of the farm. The spanwise entrainment $\Delta \mathcal{A}_{uv,y}$ is negligible in the first $10$ km of the wake, but becomes the dominant term contributing to the wake recovery in the far-wake region. However, flow entrainment along the spanwise direction is far less efficient than the vertical transfer of momentum in the wake recovery process. As a consequence, the velocity deficit shows a very slow recovery rate. Similarly to \cite{Bastankhah2024}, we note a negligible contribution of the Coriolis term $\mathcal{C}_x$, together with the spanwise turbulent transport of momentum. The background pressure gradient term $\mathcal{P}_\infty$ brings a positive contribution, although it speeds up the flow throughout the whole domain. Finally, we group terms that show an oscillatory behaviour into $\mathcal{S}=\Delta \mathcal{A}_{uu,x} + \Delta \mathcal{A}_{uw,z} + \mathcal{P}_x$, which is the sum of mean advection along the streamwise and vertical direction and pressure perturbation. The contribution of each term is shown in Figure \ref{fig:x_mom_all_H300}(b). The strong oscillatory behaviour is due to the presence of trapped waves, which considerably alter the flow behaviour within the ABL when $H$ is low. In fact, the gravity-wave horizontal wavelength predicted using linear theory measures about $4.5$~km (see \cite{Vosper2004,Sachsperger2015}), which is in line with the period of the oscillations shown in Figure~\ref{fig:x_mom_all_H300}(b). We note that this mechanism in not visible in the results of \cite{Bastankhah2024} as they fixed the capping-inversion height. Here, we can also observe that $\Delta \mathcal{A}_{uu,x}$ is out-of-phase with respect to $\Delta \mathcal{A}_{uw,z}$, while the latter is in-phase with $\mathcal{P}_x$. This behaviour can be explained through continuity. In fact, as the flow accelerates ($\Delta \mathcal{A}_{uu,x}$ negative) it also moves downward ($\Delta \mathcal{A}_{uw,z}$ positive), and vice versa. Moreover, a positive $\mathcal{P}_x$ value corresponds to a favourable pressure gradient, which causes the flow to locally accelerate, therefore generating this out-of-phase behaviour with $\Delta \mathcal{A}_{uu,x}$. The fact that both $\Delta \mathcal{A}_{uu,x}$ and $\Delta \mathcal{A}_{uw,z}$ do not attain a zero value in the far wake region underlines that the wake does not fully recover. 

\begin{figure}
	\includegraphics[width=1\textwidth]{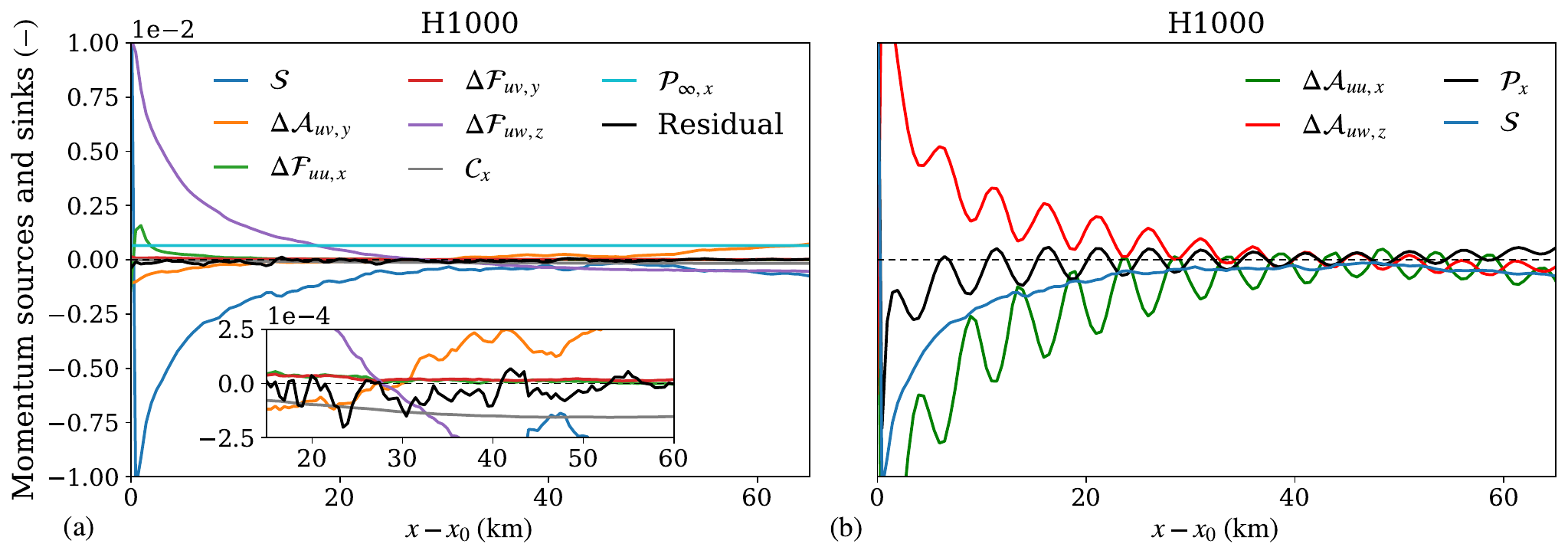} 
	\caption{Streamwise variation of momentum sources and sinks for case H1000 normalized by the time averaged total wind-farm thrust force and the control volume width. Panel (a) shows all terms of Equation \ref{eq:x_mom} including $\mathcal{S}=\Delta \mathcal{A}_{uu,x} + \Delta \mathcal{A}_{uw,z} + \mathcal{P}_x$ while panel (b) shows the contribution of each added in the $\mathcal{S}$ term. We note that only the region downwind of the farm is considered, so that the term $\mathcal{F}_{t,x}$ has no contribution. Moreover, the $x$-axis is rescaled with $x_0=30$ km which denotes the beginning of the wake region.}
	\label{fig:x_mom_all_H1000}
\end{figure}

The same analysis for case H1000 is shown in Figure \ref{fig:x_mom_all_H1000}. The main difference from case H300 is the strong vertical turbulent entrainment of momentum in the first~$20$ km of the wake, which causes the velocity deficit to rapidly recover. Since the wake is partially recovered after $20$~km, the advection of spanwise momentum by the streamwise velocity remains negligible. Similarly to case H300, $\Delta \mathcal{F}_{uu,x}$, $\Delta \mathcal{F}_{uv,y}$ and $\mathcal{C}_x$ have negligible contributions. Figure \ref{fig:x_mom_all_H1000}(b) illustrates the components of the $\mathcal{S}$ term. Here, we can observe that the flow is accelerating and moving downward (i.e. negative $\Delta \mathcal{A}_{uu,x}$ and positive $\Delta \mathcal{A}_{uw,z}$). Similarly to \cite{Bastankhah2024}, we note that the strong flow acceleration is attained roughly where the vertical turbulent transport of momentum is also high, resulting in a strong correlation between the terms $\Delta \mathcal{A}_{uu,x}$ and $\Delta \mathcal{F}_{uw,z}$. Moreover, the H1000 case has the same oscillatory frequency than case H300. This is due to the fact that the horizontal trapped-wave wavelength depends upon the capping-inversion strength and free-atmosphere lapse rate, which are equal for the two cases. We also observe the same in-phase and out-of-phase behaviour, although the magnitude of the oscillations caused by interfacial gravity waves is smaller than case H300 due to the higher distance between the turbine region and the capping-inversion height. Moreover, both $\Delta \mathcal{A}_{uu,x}$ and $\Delta \mathcal{A}_{uw,z}$ tend to zero at the end of the wake region, denoting that the wake has mostly recovered.

\begin{figure}
	\centering
	\includegraphics[width=0.55\textwidth]{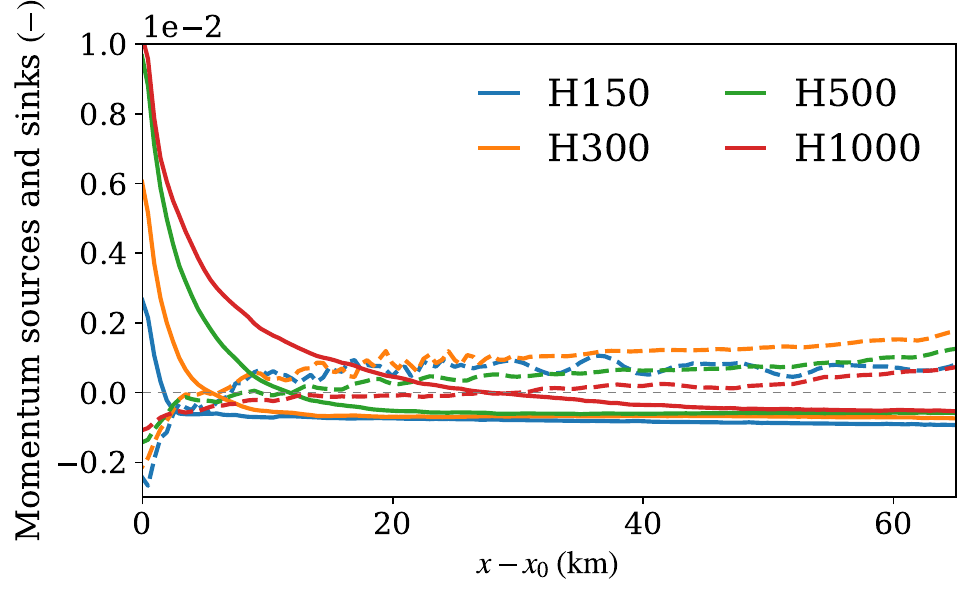} 
	
	\caption{Streamwise variation of $\Delta \mathcal{F}_{uw,z}$ (full line) and $\Delta \mathcal{A}_{uv,y}$ (dashed line) normalized by the time averaged total wind-farm thrust force of the respective case and the control volume width. We note that the $x$-axis is rescaled with $x_0=30$ km which denotes the beginning of the wake region.}
	\label{fig:x_mom_comparison}
\end{figure}

The two dominant terms that contribute to wake recovery, that is $\Delta \mathcal{F}_{uw,z}$ and $\Delta \mathcal{A}_{uv,y}$, for all cases are shown in Figure \ref{fig:x_mom_comparison}. We can see that the vertical turbulent entrainment of momentum is directly proportional to $H$. In fact, in deep boundary layer cases, the wake is mostly recovered within the first $20$ to $30$ km downwind of the farm, leaving a negligible role to the flow entrainment in the spanwise direction. Contrarily, $\Delta \mathcal{F}_{uw,z}$ dies out abruptly in shallow boundary layers, therefore causing very long wind-farm wakes, which are slowly replenished by the advection of spanwise momentum by the streamwise velocity. 

\begin{figure}
	\includegraphics[width=1\textwidth]{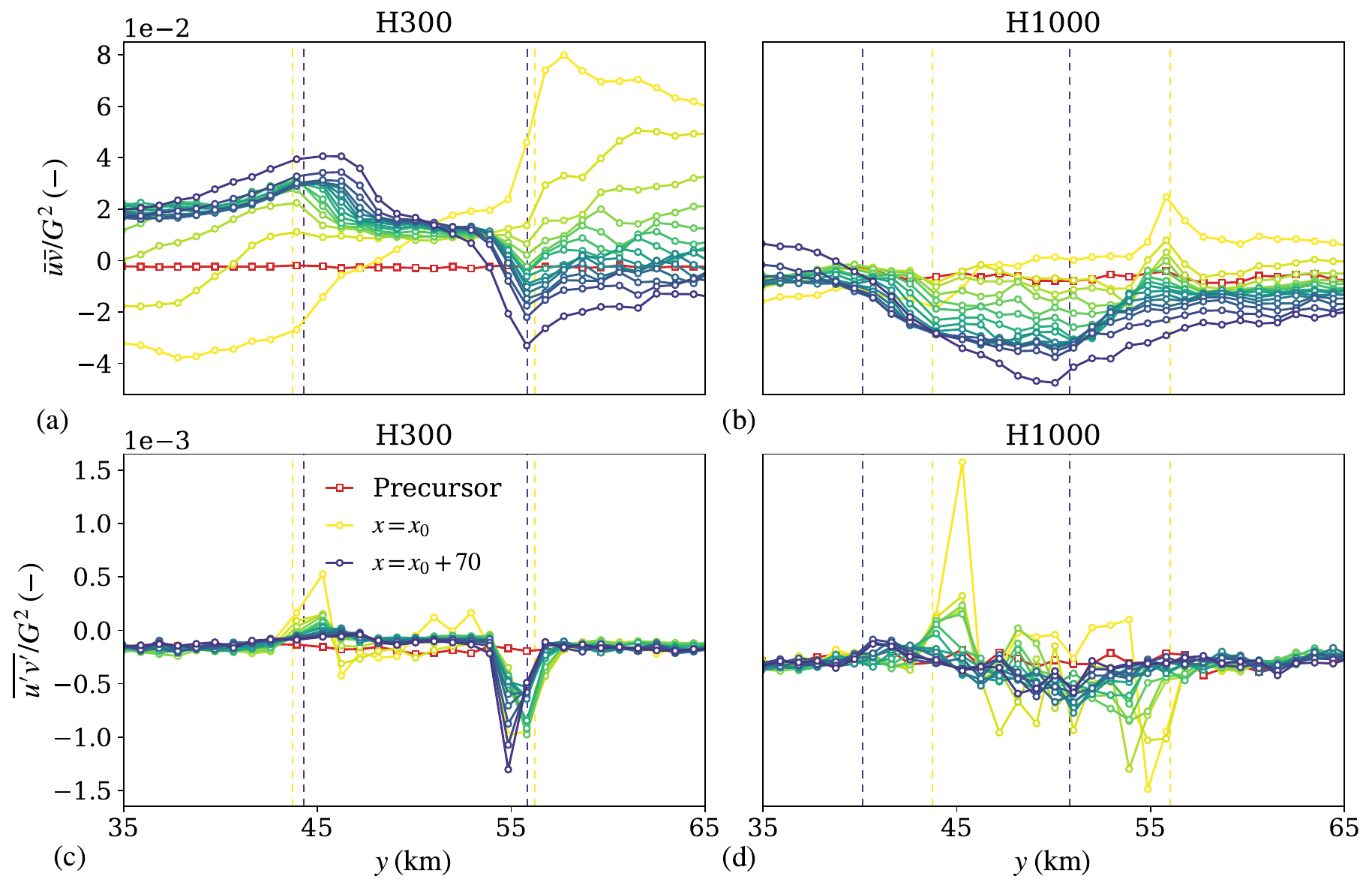}	
	\caption{Time-averaged advection of spanwise momentum by the streamwise velocity (a,b) and total spanwise turbulent transport of momentum (c,d) as a function of the spanwise direction obtained at $x=x_0$ (brightest yellow) and $x=x_0+70$ (darkest blue), with increments of $5$ km in between, for cases (a,c) H300 and (b,d) H1000. The vertical yellow and blue lines denote the location of the wake edges at $x=x_0$ and $x=x_0+70$, respectively, where $x_0=30$ km indicates the beginning of the wake region.}
	\label{fig:y_profiles_varx_uv}
\end{figure}

To further investigate the spanwise momentum entrainment induced by the mean flow and turbulence, we analyse the quantities $\bar{u} \bar{v}$ and $\overline{u' v'}$ along the $y$ direction at hub height, taken at locations downwind of the farm (i.e. from $x=x_0$ to $x=x_0+70$ with increments of $5$~km). Figure~\ref{fig:y_profiles_varx_uv}(a,b) displays the spanwise entrainment of mean flow for cases H300 and H1000, respectively. In the shallow boundary-layer case, we observe a high value of $\bar{u} \bar{v}$ at the location of the wake right edge (i.e. $y\approx 45$ km), which further increases along the downstream direction. Within the wake region, we observe a constant and positive spanwise entrainment which decreases significantly in proximity of the left edge of the wake (i.e. $y\approx 55$ km). This mechanism will be discussed in more details in Section~\ref{sec:wake_convergence}. In the H1000 case, the spanwise entrainment of mean flow attains a very similar magnitude at the wake edges, resulting in low $\Delta \mathcal{A}_{uv,y}$ values. In the far-wake region, we observe a higher spanwise entrainment of mean flow on the left edge. However, the wake deficit is almost fully recovered at these streamwise locations. Next, Figure~\ref{fig:y_profiles_varx_uv}(b,c) illustrates the spanwise turbulent transport of momentum for cases H300 and H1000. In the shallow boundary-layer case, this term is close to zero within the wake, attaining higher values in proximity of the left edge of the wake, where a steep spanwise gradient of streamwise velocity takes place -- see Figure~\ref{fig:fitting_model}(b). Since the latter increases in magnitude along the streamwise direction, we observe the peak in $\overline{u' v'}$ increasing as well. A similar behaviour is observed in the H1000 case, although higher fluctuations of $\overline{u' v'}$ are attained within the wake region. However, the role of the spanwise turbulent transport of momentum in the wake recovery process remains negligible since this term is about two orders of magnitude smaller than $\bar{u} \bar{v}$.

\subsection{Flow behaviour in and around the wake region}\label{sec:wake_convergence}
\begin{landscape}
\begin{figure}
	\includegraphics[width=1.57\textwidth]{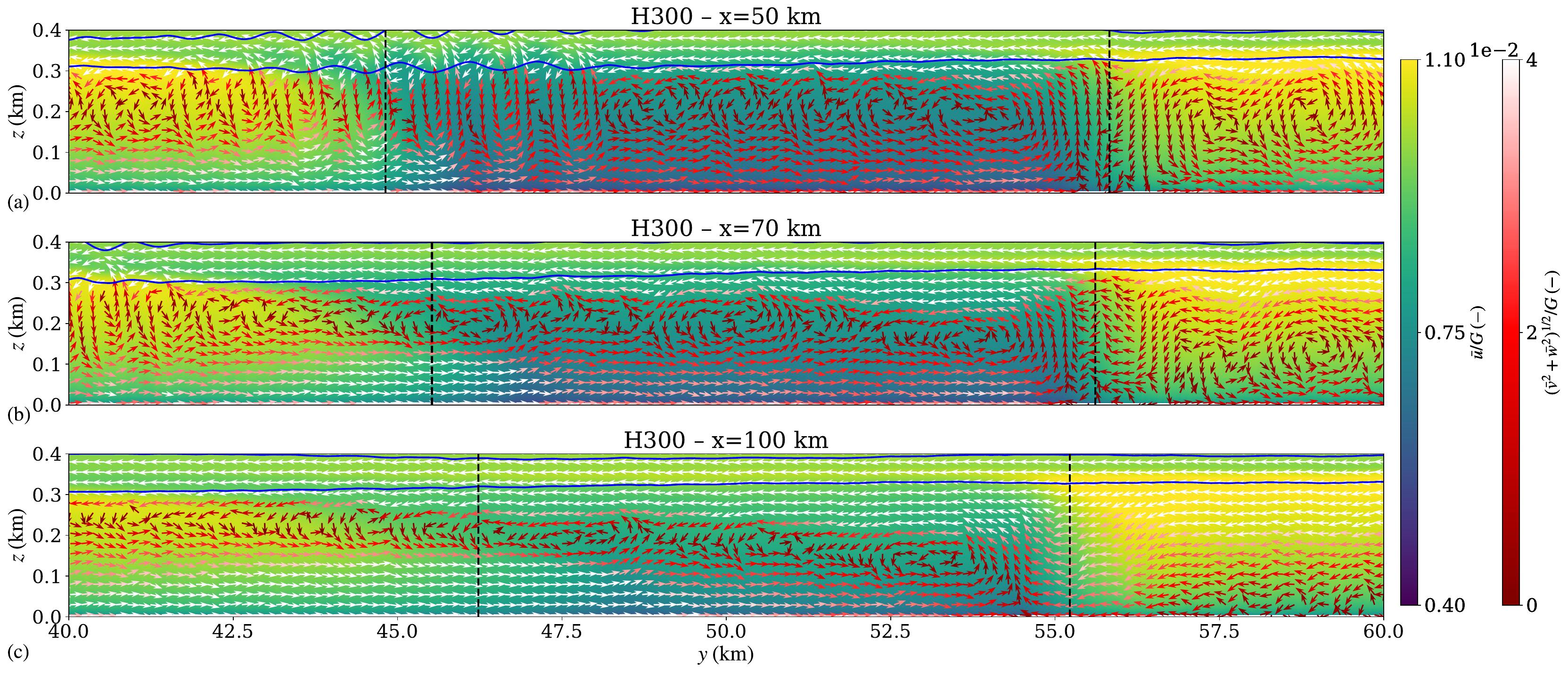} 
		
	\caption{Contours of the time-averaged streamwise velocity in an $y$--$z$ plane focused on the wake region and its sides taken at (a) $x=50$ km, (b) $x=70$ km and (c) $x=100$ km for case H300. The arrows represent the in-plane velocity vector given by the spanwise and vertical velocity components. We note that the vector is normalized with its magnitude $(\bar{v}^2+\bar{w}^2)^{1/2}$, so that all arrows have the same length. The blue lines represent the bottom and top of the inversion layer computed with the \cite{Rampanelli2004} model. Finally, the vertical black dashed lines denote the right and left wake edges predicted by the fitting model shown in Equation \ref{eq:fitting_model}.}
	\label{fig:yz_slices_H300}
\end{figure}
\end{landscape}

In this section, we investigate the flow behaviour in the wake region and at its sides. To do so, we illustrate $y$--$z$ planes of streamwise velocity taken at $x=50$, $70$ and $100$~km for case H300 in Figure~\ref{fig:yz_slices_H300}. Moreover, we add vectors which denote the flow direction based on the spanwise and vertical velocity components while their colors indicate the in-plane velocity magnitude (i.e. $(\bar{v}^2+\bar{w}^2)^{1/2}$). Further, we remark that due to the convention used, the right side of the wake corresponds to the left side of the figure, and vice versa. 

At $x=50$ km (i.e. $20$ km downstream of the farm), Figure~\ref{fig:yz_slices_H300}(a) shows a strong streamwise velocity deficit, which spreads over the whole ABL height. Within the wake and on its right side, the flow direction follows the wind veer profile of the background flow (i.e. a positive and negative spanwise velocity below and above turbine-hub height). However, the zero flow-angle angle height is shifted upward, close to turbine-tip height. This is a consequence of the reduced Coriolis force and absence of vertical mixing, which causes the flow to turn towards the pressure gradient. Moreover, we observe the presence of counterclockwise flow recirculation within the wake region, induced by the trapped waves which also corrugate the base and top of the capping inversion. The flow near the left edge of the wake behaves differently. Here, we observe a counterclockwise motion of the wake, with a positive spanwise component below tip height, an upward motion at the wake edge and a negative spanwise component near the base of the capping inversion. As a result, the spanwise velocity component attains values close to zero, significantly limiting the entrainment of mean flow at this edge (i.e. at $y\approx 56$~km). On the left side, we observe the presence of trapped waves which leads to a counterclockwise flow motion.

Figure~\ref{fig:yz_slices_H300}(b,c) shows the flow behaviour at $x=70$ and $x=100$~km, respectively. As we move downstream, the streamwise velocity component recover at a slow rate along the downstream direction while the wake gradually narrows. The zero flow-angle height moves downward towards turbine-hub height, with the wind veer profile converging towards the one of the background flow within the wake and on its right side. However, the counterclockwise motion at the left edge of the wake does not die out and is still clearly visible $50$~km downstream of the farm. Contrarily, the trapped waves at the left side of the wake are gradually dampened along the streamwise direction, so that the counterclockwise flow rotation at the left side of the wake vanishes at $x=100$ km and the flow assumes a negative spanwise component at each height. Finally, Figure~\ref{fig:yz_slices_H300}(b,c) also illustrates how the strong wind veer attained on the right edge of the wake causes the streamwise velocity deficit to spread horizontally.

\begin{landscape}
	\begin{figure}
		\includegraphics[width=1.57\textwidth]{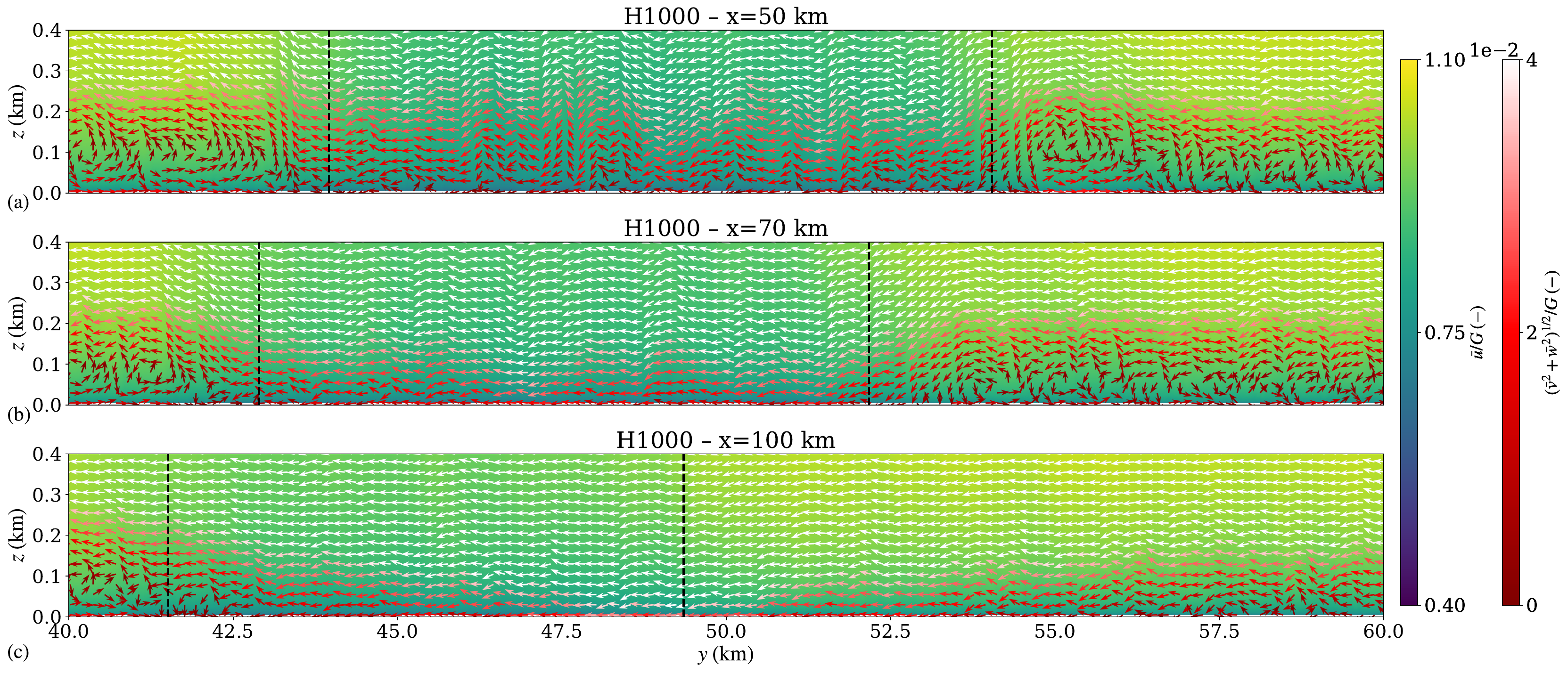} 
		
		\caption{Contours of the time-averaged streamwise velocity in an $y$--$z$ plane focused on the wake region and its sides taken at (a) $x=50$ km, (b) $x=70$ km and (c) $x=100$ km for case H1000. The arrows represent the in-plane velocity vector given by the spanwise and vertical velocity components. We note that the vector is normalized with its magnitude $(\bar{v}^2+\bar{w}^2)^{1/2}$, so that all arrows have the same length. Finally, the vertical black dashed lines denote the right and left wake edges predicted by the fitting model shown in Equation \ref{eq:fitting_model}.}
		\label{fig:yz_slices_H1000}
	\end{figure}
\end{landscape}

The same analysis for case H1000 is shown in Figure~\ref{fig:yz_slices_H1000}. Here, the flow behaviour is significantly different than the one obtained in shallow boundary layers. In fact, the strong flow mixing along the vertical direction adds clockwise-turning flow from aloft into the wake region. Consequently, Figure~\ref{fig:yz_slices_H1000}(a) illustrates that the streamwise velocity deficit recovers at a much faster rate with the spanwise velocity component being negative at each height within the wake region. Moreover, the wind veer is significantly reduced if compared to the one of the background flow. Further, we also observe a much stronger in-plane velocity magnitude compared to case H300. At the sides of the wake, the wind veer follows the same orientation of the one attained within the wake region. Moreover, flow recirculation induced by trapped waves remains very limited. In fact, the influence of trapped waves on the flow behaviour within the rotor height is very limited in deep boundary layers -- see Figure~\ref{fig:xy_slices}(l). Further downstream, Figure~\ref{fig:yz_slices_H1000}(b,c) shows that the streamwise velocity deficit is mostly recovered and the wake deflects clockwise. Moreover, the streamwise velocity deficit does not spread along the horizontal direction as a consequence of the reduced wind veer. 

\section{Conclusions}\label{sec:conclusions}
The aim of this work is to bring new physical insights into the wake behaviour of large-scale wind farms operating in CNBLs with various capping-inversion heights. To this end, we performed four simulations of a single wind farm using SP-Wind, an LES solver developed at KU Leuven. The main domain was driven by turbulent fully developed statistically steady flow fields obtained in precursor simulations. Moreover, the solver periodicity in the streamwise direction was broken using a wave-free fringe region technique while at the top of the domain an RDL was placed to damp gravity waves. A large numerical domain with length, width and height of $110 \times 100 \times 25$~km$^3$ was used to allow the wake to fully develop and to minimize artificial effects due to the domain boundaries.

In shallow boundary layers, such as cases H150 and H300, the vicinity of the capping inversion to the turbine-tip height limits the wake development in the vertical direction. Moreover, the capping inversion behaves as a pliant surface, limiting flow entrainment from the free atmosphere. Consequently, these cases show strong velocity deficits in the farm wake and a very slow recovery rate. The strong flow deceleration attained in these two cases decreases the Coriolis force magnitude. Therefore, the wake turns towards the direction of the background pressure gradient, that is an anticlockwise flow rotation in the Norther Hemisphere. In cases H500 and H1000, the increase in the $H/z_\mathrm{tip}$ ratio allows the formation of a high speed channel between the tip height and the capping-inversion base which enhances vertical turbulent entrainment of momentum into the wake, therefore giving rise to a lower velocity deficit and a faster recovery rate. The vertical entrainment of momentum adds clockwise-turning flow from aloft into the wake region. Therefore, a clockwise wake deflection is observed, which is dictated by the wind veer present in the background flow. This also means that, in flows with small wind veer, the wind-farm wake would turn anticlockwise. Despite a distance of roughly $75$~km from the last-row turbine to the fringe region, in none of the cases the wake was fully replenished, with velocity deficits with respect to the inflow up to $25$\% measured $70$ km downstream of the farm in shallow boundary layers. Similarly to previous studies, we observed a much faster recovery in terms of TKE, which drops below the background value within the first $10$ to $20$ km into the wake.

Defining a wake region together with estimating wake properties is non-trivial. Therefore, we proposed a simple fitting model for the spanwise velocity magnitude profiles at streamwise locations downwind of the farm. The latter allowed us to evaluate wake properties such as the wake strength, width and deflection. We observed that the wake narrows in all cases along the streamwise direction, with a wake-to-farm width ratio going from about $1.2$ to $1$ over a distance of $70$ km. The wake center predicted by the fitting model allowed us to clearly observe how the capping-inversion height influences the wake deflection, generating a clear distinction between shallow and deep boundary layers. Moreover, we observed a very fast recovery rate in the first $5$ km into the wake, which slows down considerably further downwind. Furthermore, the model allowed us to observe that the wake edges and center do not vary across the rotor height, besides for case H150 where a significant clockwise rotation at tip height takes place. This effect is caused by the strong wind veer attained within the capping inversion, which is very near to tip height in this case.

To highlight the dominant wake recovery mechanisms, we performed a streamwise momentum budget analysis. We found that the two dominant terms that contribute to wake recovery are the vertical turbulent entrainment of momentum and the advection of streamwise momentum by the spanwise velocity. However, their contribution differs depending on the capping-inversion height. In shallow boundary layers, the close vicinity of the capping inversion to turbine tip-height dampens the vertical turbulent entrainment of momentum, which drops abruptly in the first $5$ km into the wake. Consequently, the wake is mostly recovered by spanwise mean flow entrainment, which is rather inefficient therefore causing very long wakes with strong deficits. Further, we noticed that the spanwise entrainment of mean momentum mostly occurs on the right edge of the wake. In fact, a counterclockwise flow motion near the left edge of the wake, which persists at each location downstream of the farm, provokes strong local spanwise gradients of streamwise velocity and significantly limits the spanwise entrainment of mean flow in this region. In deep boundary layers, the wake is mostly recovered by vertical turbulent entrainment of momentum and spanwise entrainment has a negligible contribution. In such cases, the wind veer profiles in the wake region and at its sides are very similar and the effects of trapped waves on the flow behaviour are limited. Consequently, we do not observe regions of flow recirculation and convergence at the wake edges.

Finally, we conducted a study on the sensitivity of the numerical solution to the grid resolution and domain width using the H300 precursor flow fields to drive the simulations. Interestingly, we found that the error caused by coarsening the grid resolution is an order of magnitude smaller than the error introduced by using a domain with too low $L_y/L_y^f$ ratio. Moreover, we observed that doubling the grid cell size in the streamwise and spanwise direction does not alter the wake properties. However, a domain with ratio $L_y/L_y^f$ of 3.19 displays a narrower farm wake with a faster recovery rate than the one observed in a domain with $L_y/L_y^f=9.57$. Therefore, particularly when simulating shallow boundary layers, we recommend to use a sufficiently wide domain at the expenses of using a coarser numerical grid.  

In the current work, we solely focused on the wake development of a single farm. However, farms are typically clustered together nowadays. Therefore, a natural extension of this work consists in considering the wake behaviour of a cluster of farms. This would also allow to better investigate farm--farm interactions, together with their effects on power losses and loads. Furthermore, the study assumes the presence of statistically steady state conditions, which rarely occur in a real scenario. Therefore, we are planning to drive our simulations using mesoscale forcing to also account for the effects of the natural variability in the inflow conditions on the wake development. This study limited its scope to CNBLs. However, we are aware that the wake dynamic is strongly influenced by stability effects within the ABL. Therefore, future research should also focus on investigating farms wake strength and deflection in stable and convective boundary layers. Finally, analytical wake models are known to over predict the wake recovery rate. Therefore, the set of simulations analyzed in this work can be used as a benchmark for future development and calibration of faster engineering models.\\

\noindent
\textbf{Acknowledgements.} \\
The authors thank Steven Vandenbrande for HPC technical support. \\
\\
\noindent
\textbf{Funding.} \\
The authors acknowledge support from the Research Foundation Flanders (FWO, Grant No. G0B1518N), from the project FREEWIND, funded by the Energy Transition Fund of the Belgian Federal Public Service for Economy, SMEs, and Energy (FOD Economie, K.M.O., Middenstand en Energie) and from the European Union Horizon Europe Framework programme (HORIZON-CL5-2021-D3-03-04) under grant agreement no. 101084205. The computational resources and services in this work were provided by the VSC (Flemish Supercomputer Center), funded by the Research Foundation Flanders (FWO) and the Flemish Government department EWI. \\
\\
\noindent
\textbf{Declaration of interests.}\\
The authors report no conflict of interest.\\
\\
\noindent
\textbf{Data availability statement.}\\
The full dataset underlying the analysis in the current paper, including, e.g, turbine statistics, three-dimensional time-averaged flow fields of first- and second-order statistics of precursor and main domains, as well as some post-processing example scripts are available open source as a KU Leuven RDR dataset: https://doi.org/10.48804/LRSENQ \citep{Lanzilao_dataset2024}. \\
\\
\noindent
\textbf{Author contributions.}\\
L.L. and J.M. jointly set up the simulation studies and wrote the manuscript. L.L. performed code implementations and carried out the simulations. J.M. supervised the research and was responsible for acquisition of funding.\\

\appendix

\section{Influence of the horizontal grid resolution and domain width on the numerical results}\label{app:domain_sensitivity}
The simulations presented in this article were initially performed on a domain with width of $30$ km and a finer horizontal grid resolution. However, \cite{Lanzilao2024} have shown that the domain width can have major influences on the results in simulations of wind farms operating in CNBLs. In particular, they recommended a $L_y/L_y^{f}$ ratio of at least 6 in presence of shallow boundary layers. Therefore, we have re-run all cases on a wider domain. Moreover, we have used a coarser grid resolution, to keep the computational cost manageable. This new set of simulations is the one analyzed in the main body of the article. Here, we use the original batch of simulations to illustrate the influence of the domain width and grid resolution on the wake behaviour. We note that the comparison is only made with cases driven by the H300 precursor flow fields. For other set-up parameters, we refer the reader to Section \ref{sec:methodology}.

Two different numerical grids are analyzed. \cite{Lanzilao2022,Lanzilao2022b,Lanzilao2024} have used a horizontal grid with cell sizes of $\Delta x/2$ and $\Delta y/2$, where $\Delta x$ and $\Delta y$ measures 62.5 and 43.48 m, respectively. We define this as the fine grid, which we use to run a simulation on a numerical domain with length and width of $90$ and $30$ km, respectively. This simulation will be denoted as the Ly-30-fine case. The second grid, which we define as coarse, uses a streamwise and spanwise grid size corresponding to $\Delta x$ and $\Delta y$. Using this grid, we perform three additional simulations where the domain length is fixed to $90$ km while the width assumes values of $30$, $60$ and $90$ km. The latter cases will be denoted as Ly-30-coarse, Ly-60-coarse and Ly-90-coarse, respectively. Here, it is necessary to use a coarser grid to make the simulations affordable, given the very high number of DOF. The vertical grid and domain dimension correspond to the ones described in Section \ref{sec:numerical_setup} and are equal for all cases. The error caused by the coarsening of the horizontal grid can be estimated by comparing the Ly-30-fine and Ly-30-coarse cases, as the only difference here is streamwise and spanwise dimension of the grid cell. The differences caused by the domain width can be estimated by comparing case Ly-30-coarse versus cases $L_y$-60-coarse and $L_y$-90-coarse, since the grid resolution is equal for all cases. The outcome of this analysis will allow us to chose a domain size and a numerical grid suitable for these type of simulations. We note that the cases considered in this section are listed in Table~\ref{table:domsens_setup}, where additional information about the cases set-up is reported. 

\begin{table}
	\def~{\hphantom{0}}
	\begin{center}
		\begin{adjustbox}{max width=\textwidth}
			\begin{tabular}{cccccccccc}
				\textbf{Cases}  & $\boldsymbol{L_x}$ \textbf{(km)} & $\boldsymbol{L_y}$ \textbf{(km)} & $\boldsymbol{L_z}$ \textbf{(km)} & $\boldsymbol{L_\mathrm{wake}/L_x^f}$ \textbf{(--)} & $\boldsymbol{L_y/L_y^f}$ \textbf{(--)} & $\boldsymbol{N_x}$ \textbf{(--)} & $\boldsymbol{N_y}$ \textbf{(--)} & $\boldsymbol{N_z}$ \textbf{(--)} & \textbf{DOF (--)}\\[7pt]
				Ly-30-fine      & 90 & 30 & 25 & 3.68 & 3.19 & 2880 & 1380 & 490 & $8.66$ $\times 10^9$\\
				Ly-30-coarse    & 90 & 30 & 25 & 3.68 & 3.19 & 1440 & 690 & 490 & $2.16$ $\times 10^9$\\ 
				Ly-60-coarse    & 90 & 60 & 25 & 3.68 & 6.38 & 1440 & 1380 & 490 & $4.33$ $\times 10^9$\\ 
				Ly-90-coarse    & 90 & 90 & 25 & 3.68 & 9.57 & 1440 & 2070 & 490 & $6.49$ $\times 10^9$\\
				Selected & 110 & 100 & 25 & 5.03 & 10.64 & 1760 & 2300 & 490 & $8.66$ $\times 10^9$\\[1pt]
			\end{tabular}
		\end{adjustbox}
	\end{center}
	\caption{Overview of the numerical domains used to perform the sensitivity study on the horizontal grid resolution and domain width. The parameters $L_x$, $L_y$ and $L_z$ denote the streamwise, spanwise and vertical domain dimensions while $N_x$, $N_y$ and $N_z$ denote the number of grid points used along the respective directions. Moreover, $L_\mathrm{wake}$ denotes the fetch of domain between the last-row turbine location and the start of the fringe region while $L_x^f$ and $L_y^f$ represent the wind-farm length and width, respectively. The last column reports the number of DOF comprehensive of both precursor and main domains. We note that each case is driven by the H300 precursor flow fields. Finally, the last row reports the selected domain used to perform the sensitivity study to the atmospheric state.}
	\label{table:domsens_setup}
\end{table}

\begin{figure}
	\centering
	\includegraphics[width=0.49\textwidth]{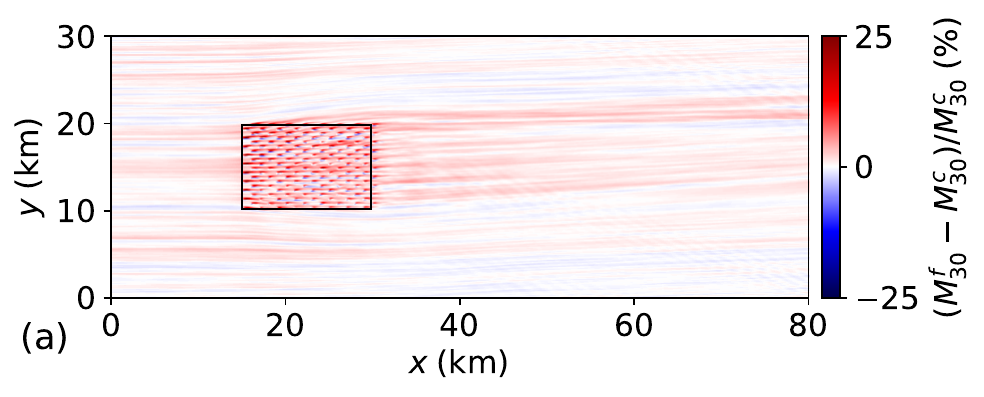} 
	\includegraphics[width=0.49\textwidth]{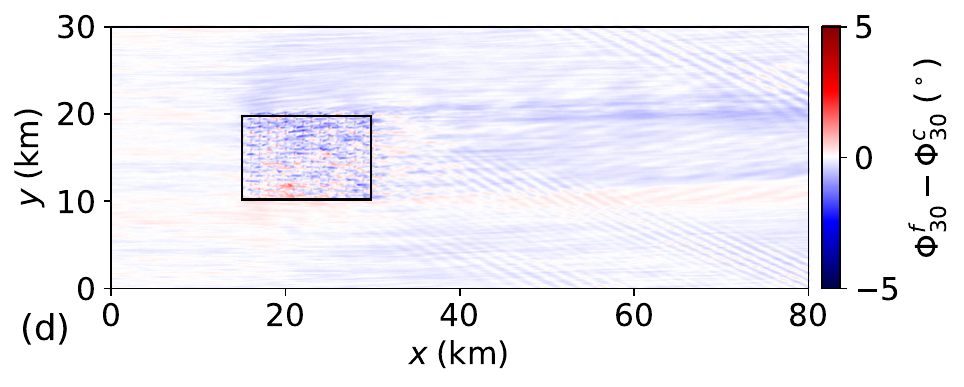}
	
	\includegraphics[width=0.49\textwidth]{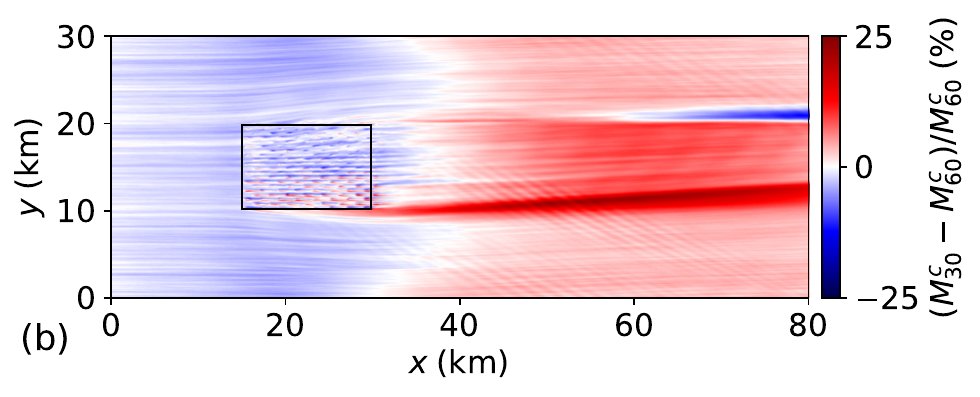}
	\includegraphics[width=0.49\textwidth]{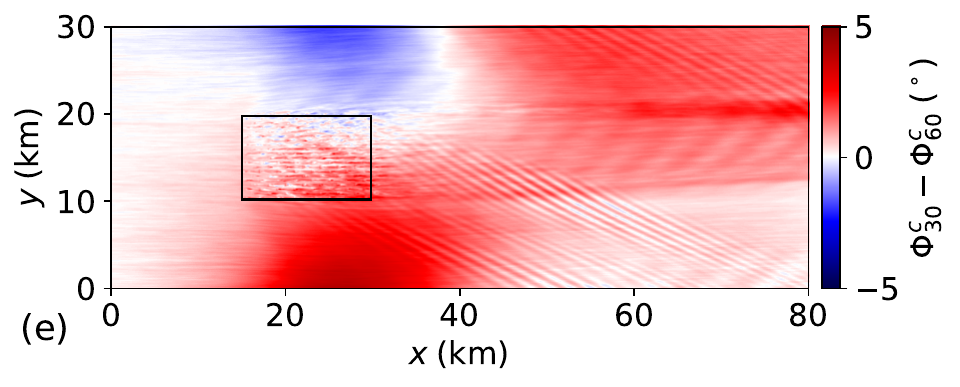}
	
	\includegraphics[width=0.49\textwidth]{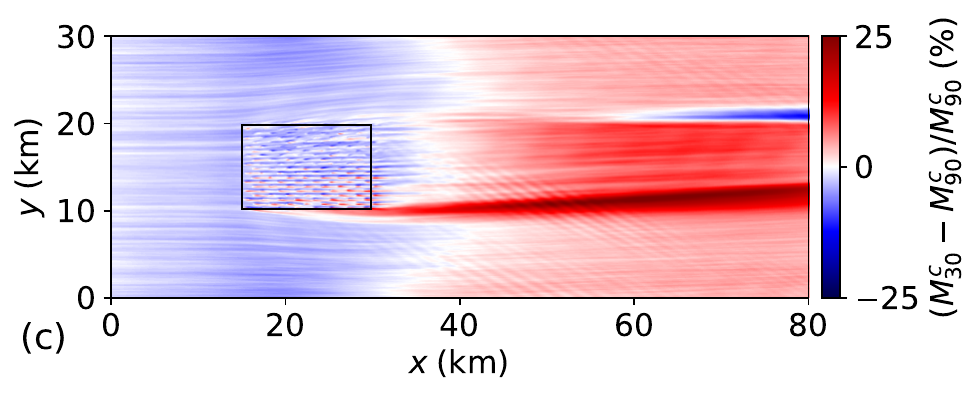}
	\includegraphics[width=0.49\textwidth]{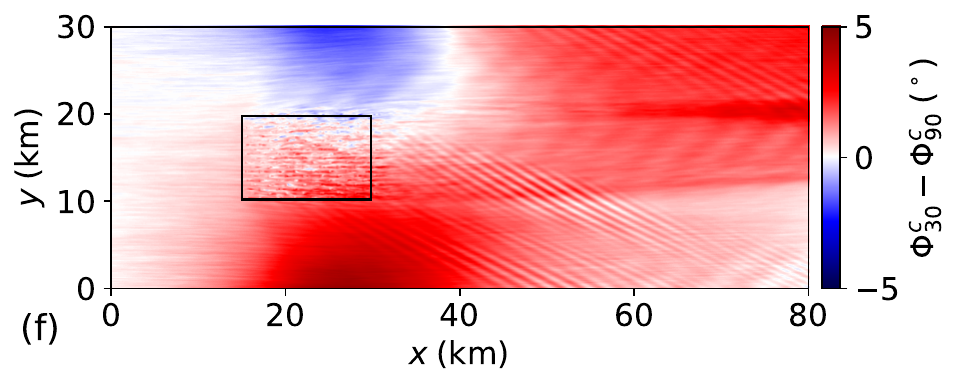}
	
	\caption{(a-c) Relative error in terms of velocity magnitude and (d-f) flow angle in an $x$--$y$ plane taken at hub height. Panels (a-d) highlight differences due to the grid resolution while panels (b,c,e,f) show differences due to the domain width. The location of the wind farm is indicated by the black rectangle.}
	\label{fig:domsens_fields}
\end{figure}

We start by discussing the impact of the horizontal grid resolution on the numerical simulation. First of all, we remark that the results in the precursor domain are not affected by the change in grid resolution. This is visible in Figure \ref{fig:precursor_results}. Therefore, all differences observed here are solely due to the flow dynamics in the main domain. Figure \ref{fig:domsens_fields}(a,b) shows the relative error in terms of velocity magnitude and the difference in terms of flow angle between cases Ly-30-fine and Ly-30-coarse. We observe that the velocity within the farm area at hub height is on averaged about 3 to 5\% higher when using the fine grid. The wake recovery appears to be slightly slower when using the coarse grid, although differences tend to zero towards the end of the domain. Overall, we observe that the major discrepancies are attained within the farm and on the left edge of the wake. This is expected since those are regions of strong velocity gradients. In terms of flow angle, we observe differences up to $1^\circ$, which mostly occur in the far-wake region. 

Next, we discuss the impact of the domain width on the results. Figure \ref{fig:domsens_fields}(c,d) shows the comparison in terms of velocity magnitude and flow angle between cases Ly-30-coarse and Ly-60-coarse, where the ratio $L_y/L_y^f$ varies from 3.19 to 6.38. It is easy to observe how the impact of the domain width on the numerical results is at least one order of magnitude higher than the one caused by the grid resolution. For instance, similarly to \cite{Lanzilao2024}, we observe that a smaller domain overestimates flow blockage, with the velocity being in the order of 10\% lower in front of the farm. Moreover, a narrower domain enhances the wake recovery rate and its anticlockwise deflection. The same phenomena are observed in Figure \ref{fig:domsens_fields}(e,f) where we compare the Ly-30-coarse and Ly-90-coarse cases. Here, velocity differences in the far wake region reaches value above 25\% while differences in flow angle are up to $5^\circ$.  

\begin{figure}
	\includegraphics[width=0.5\textwidth]{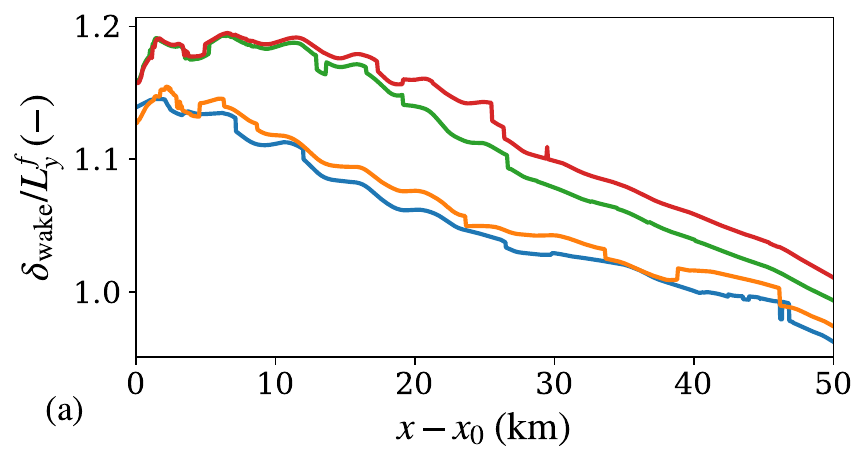} 
	\includegraphics[width=0.5\textwidth]{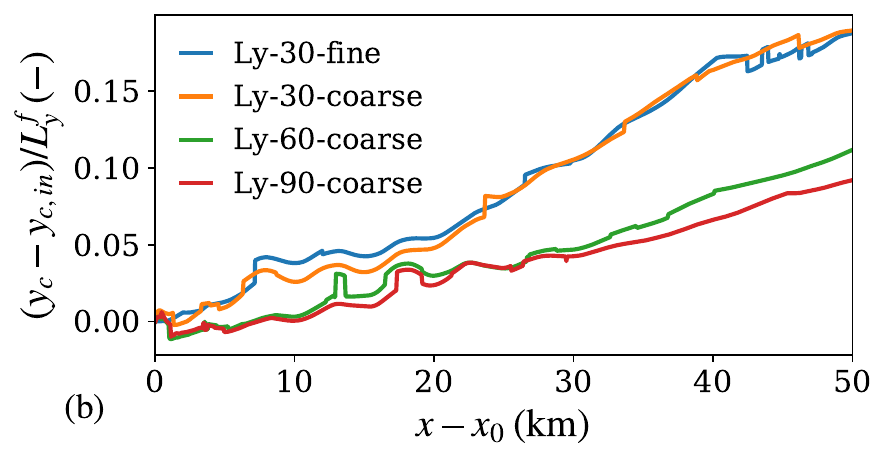} 
	\includegraphics[width=0.5\textwidth]{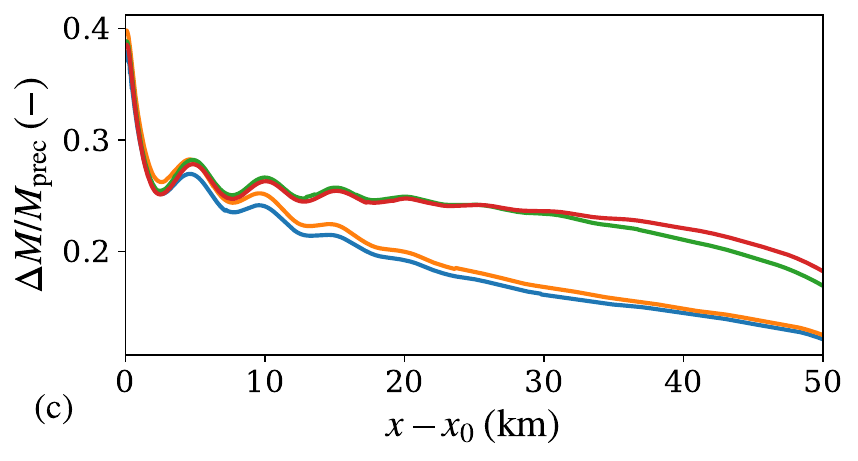} 
	\includegraphics[width=0.5\textwidth]{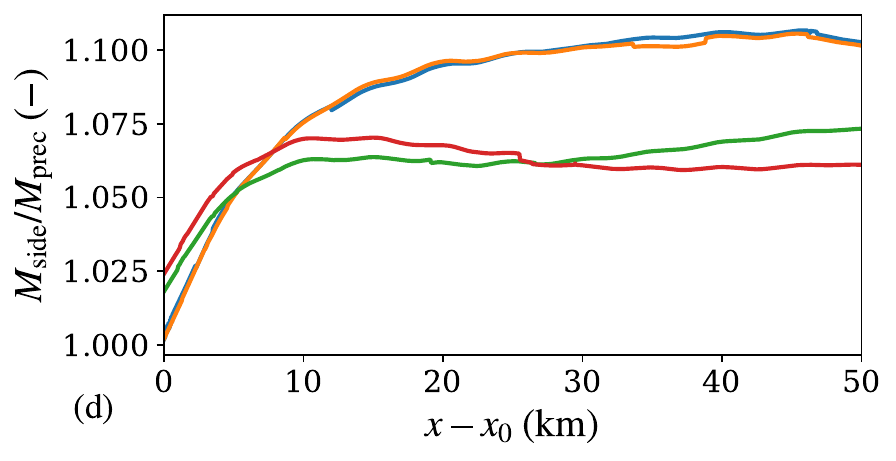} 
	
	\caption{(a,b) Wake width and wake center normalized with the farm width, (c,d) velocity deficit with respect to $M_\mathrm{prec}$ and velocity magnitude at the sides of the wake normalized with the velocity magnitude obtained in the precursor domain at hub height. All quantities are predicted by the fitting model shown in Equation~\ref{eq:fitting_model}. Moreover, the $x$-axis is rescaled with $x_0=30$ km which denotes the beginning of the wake region.}
	\label{fig:domsens_wake_properties}
\end{figure}

We now discuss the differences in wake width, strength and deflection within the wake and at its sides by using the fitting model reported in Equation \ref{eq:fitting_model}. The results are shown in Figure \ref{fig:domsens_wake_properties}. By comparing cases Ly-30-fine and Ly-30-coarse, we can observe how the wake width and center assume very similar values. A similar conclusion can be drawn when comparing the wake velocity deficit and $M_\mathrm{side}$. Therefore, we can conclude that doubling the grid cell size in the streamwise and spanwise directions has little influence on the wake properties. In terms of domain width, differences are noticeable, particularly when comparing cases Ly-30-coarse and Ly-90-coarse. Figure \ref{fig:domsens_wake_properties}(a) shows that the wake narrows in both cases. However, $\delta_\mathrm{wake}/L_y^f$ is about 0.05 larger in the Ly-90-coarse case. Further, Figure \ref{fig:domsens_wake_properties}(b) shows that a narrow domain artificially enhances the anticlockwise deflection of the wake. Figure \ref{fig:domsens_wake_properties}(c) illustrates that the velocity deficit in the wake, here defined with respect to the precursor velocity at hub height, recovers faster in a narrower domain. The high speed channels that develops at the sides of the wake in domains with low $L_y/L_y^f$ ratios can be responsible for this effect. In fact, Figure \ref{fig:domsens_wake_properties}(d) shows a clear difference in the $M_\mathrm{side}$ profile between cases Ly-30-coarse and Ly-90-coarse. In fact, the velocity at the sides of the wake keeps increasing along the streamwise direction in a narrow domain, reaching $M_\mathrm{side}/M_\mathrm{prec}$ values up to 1.1. This phenomenon was also observed by \cite{Lanzilao2024}. In wide domains, $M_\mathrm{side}$ reaches a peak $10$ km downstream of the farm, remaining constant afterwards.

\begin{figure}
	\centering
	\includegraphics[width=0.5\textwidth]{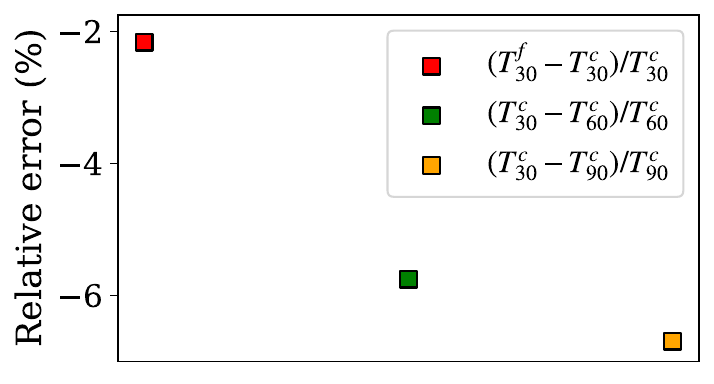}
	
	\caption{Relative error in terms of total farm thrust force for cases with different grid resolution and domain width.}
	\label{fig:domsens_ftot}
\end{figure}

The strength of the wake generated by a farm is dependent on the amount of force that the turbines exert on the flow. Therefore, we conclude this analysis by investigating how the total thrust $T$ that the farm exerts on the incoming flow depends upon the grid size and domain size. The results are reported in Figure \ref{fig:domsens_ftot}, which shows the relative error in terms of thrust force. Doubling the size of the horizontal grid cells in the $x$ and $y$ directions introduces a relative error of $2\%$ on $T$. However, tripling the domain width induces a difference of about $7\%$.

We note that the presence of periodic boundary conditions in the horizontal directions allows for alternative interpretations of these results. For instance, when the domain width is $30$~km and the $L_y/L_y^f$ ratio measures 3.19, this would correspond to an infinite row of wind farms with spacing of $L_y-L_y^f=20.6$~km. From this perspective, we can assert that the wake recovery rate in a row of wind farms is faster than that of an isolated farm, accompanied by a more pronounced wake deflection. Moreover, similarly to a single row of turbines \citep{Mctavish2015}, smaller spanwise distances between farms generate high speed channels which can enhance the power output of downstream farm. Although the simulations performed by \cite{Maas2022} show similar results, the wake behaviour in clusters of farms requires more attention and further investigation.

In conclusion, we have observed that the wake behaviour is mostly insensitive to the grid resolutions adopted here. Moreover, the results indicate that a narrow domain alter substantially the wake properties, enhancing the wake recovery and altering the wake deflection. Moreover, a domain length of $90$ km with $L_\mathrm{wake}/L_x^f=3.68$ does not suffice for attaining a full wake recovery (i.e. the wake is longer than $55$ km). As a result of this study, we fix the main domain length and width to $110$ and $100$ km, respectively. This allows us to increase $L_\mathrm{wake}/L_x^f$ up to $5.68$, while $L_y/L_y^f=10.64$, which is well above the value of 6 suggested by \cite{Lanzilao2024}. Since results are quasi-independent from the grid size, we adopt the coarse grid, which implies a $\Delta x$ and $\Delta y$ of $62.5$ and $43.48$ m, respectively, with a total of $8.66 \times 10^9$ DOF. The selected numerical set-up is reported in Table \ref{table:domsens_setup}.

\section{Influence of the capping-inversion thickness on the wake behaviour}\label{app:ci_thickness}
\begin{figure}
	\centering
	\includegraphics[width=1.\textwidth]{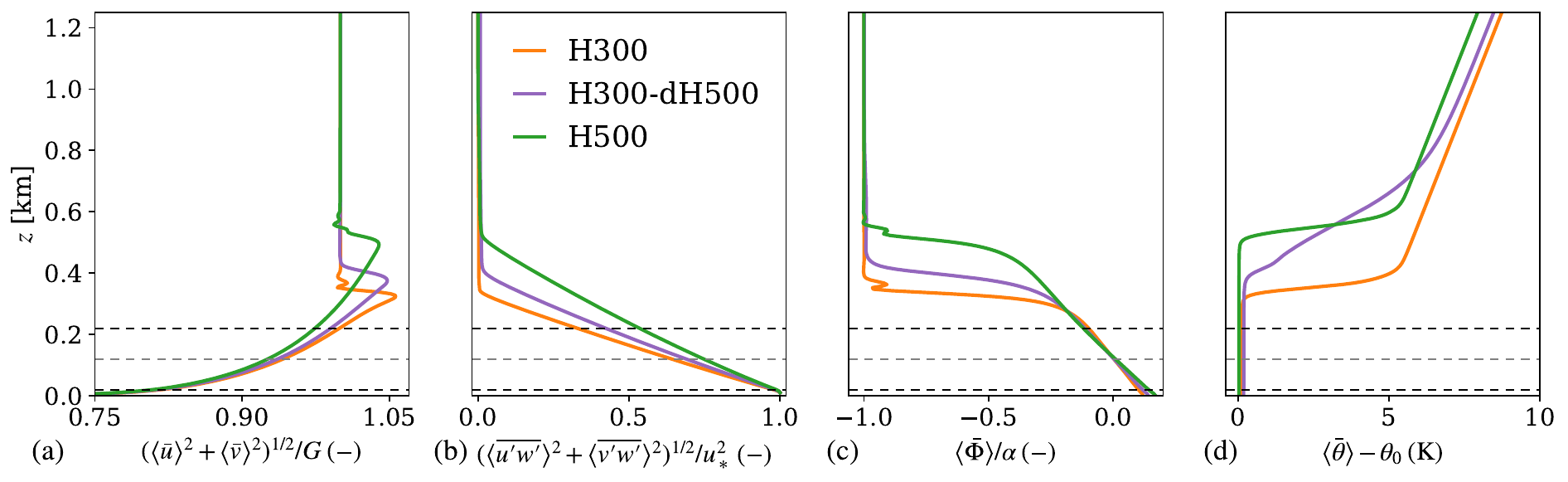}%
	\caption{Vertical profiles of (a) velocity magnitude, (b) total shear stress magnitude, (c) wind direction and (d) potential temperature averaged along the full horizontal directions and over the last $4$ h of the simulation. The continuous lines denote the profiles used in this work while the dashed lines represent the profiles obtained with a finer horizontal grid resolution with $\Delta x/2$ and $\Delta y/2$ (i.e. the profiles used by \cite{Lanzilao2024}). Finally, the grey dashed line denotes the turbine-hub height while the black dashed lines are representative of the rotor dimension. We note that the results shown here only refer to the precursor simulations.}
	\label{fig:precursor_results_dh}
\end{figure}

The aim of this appendix is to investigate the sensitivity of the wake behaviour to the capping-inversion thickness. To this end, we will compare the results obtained in cases H300 and H500, which adopt a capping-inversion thickness of $100$ m, against case H300-$\Delta$H500, that is a simulation performed using the same setup of case H300 but with a $\Delta H$ of 500~m. The time averaged profiles over the last $4$ hour of precursor simulation are displayed in Figure \ref{fig:precursor_results_dh}. Here, we observe that a thicker capping inversion allows for a bigger growth of the ABL during the spin-up phase. In fact, the capping-inversion height for cases H300 and H300-$\Delta$H500 corresponds to $319$ and $397$ m, respectively. Therefore, the differences in results between these two cases are not only due to a change in capping-inversion thickness, but also to a change in capping-inversion height.   

\begin{figure}
	\centering
	\includegraphics[width=1.\textwidth]{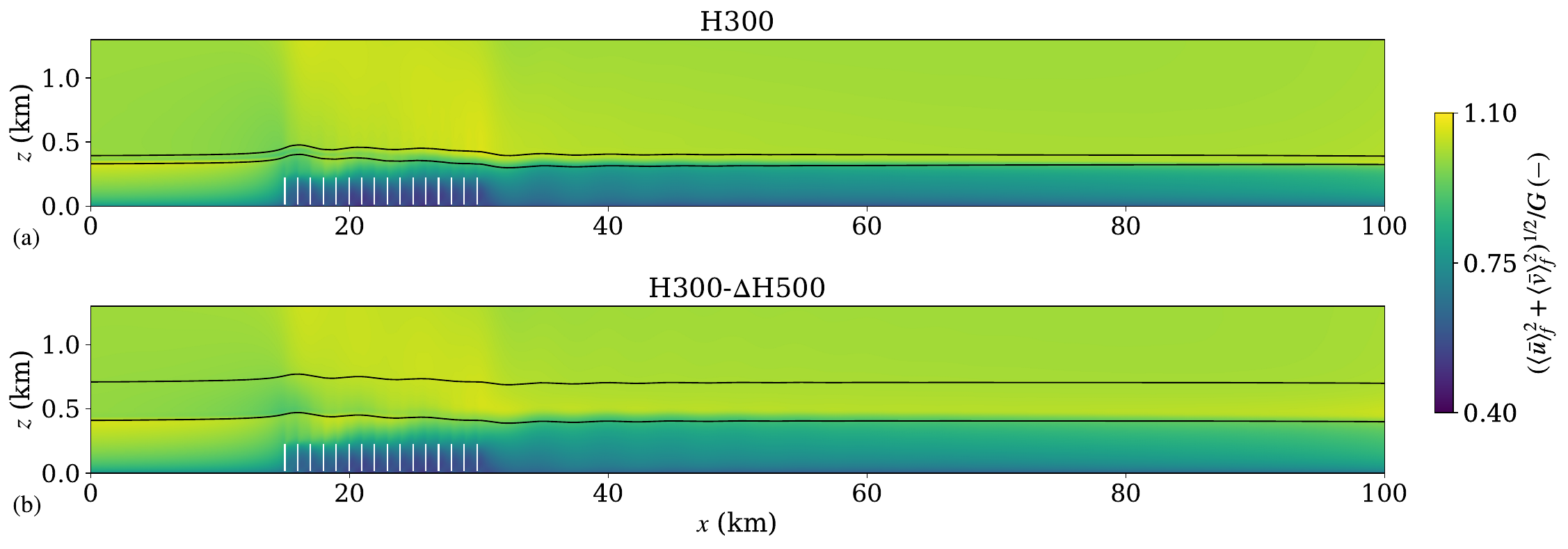}
	\caption{Contours of the time-averaged horizontal velocity magnitude in an $x$--$z$ plane further averaged along the farm width in the spanwise direction for cases (a) H150, (b) H300, (c) H500 and (d) H1000. The black lines represent the bottom and top of the inversion layer computed with the \cite{Rampanelli2004} model. Finally, the location of the turbine-rotor disks is indicated with vertical white lines.}
	\label{fig:xz_slices_dh}
\end{figure}

Figure \ref{fig:xz_slices_dh}(a,b) shows a side view of horizontal velocity magnitude averaged in the $y$ direction along the width of the farm for cases H300 and H300-$\Delta$H500, respectively. Moreover, the black lines illustrate the base and top of the inversion layer computed by fitting the LES data with the \cite{Rampanelli2004} model, which clearly illustrate the difference in $\Delta H$ between the two cases. We can also observe a difference in velocity deficit, with a stronger wake attained in case H300. Despite the capping inversion being five times thicker in case H300-$\Delta$H500, we observe that the velocity deficit does not penetrate into the inversion layer, which limits the vertical development of the wake. 

\begin{figure}
	\includegraphics[width=0.5\textwidth]{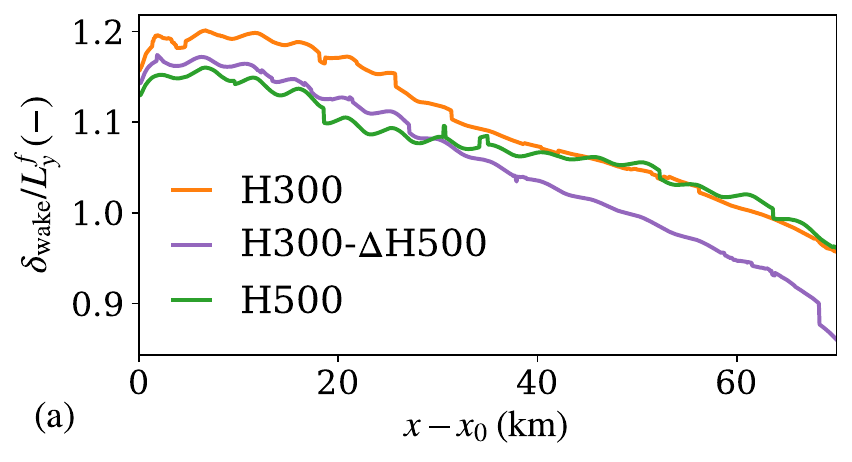} 
	\includegraphics[width=0.5\textwidth]{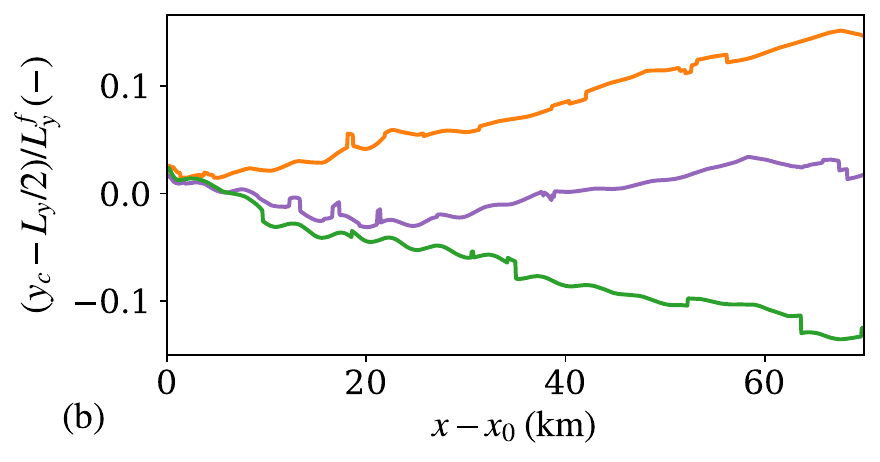} 
	\includegraphics[width=0.5\textwidth]{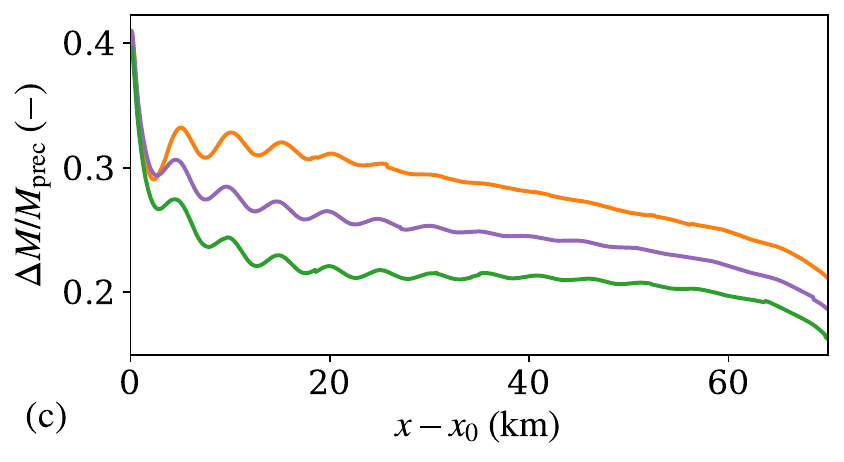} 
	\includegraphics[width=0.5\textwidth]{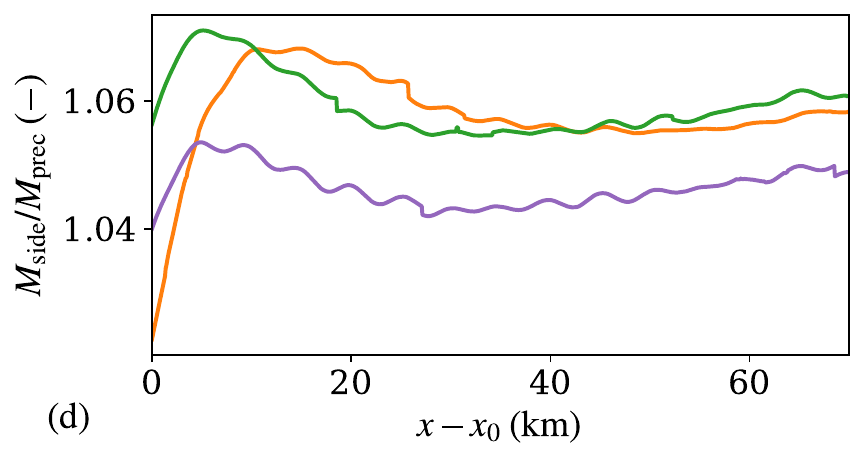} 
	
	\caption{(a,b) Wake width and wake center normalized with the farm width, (c,d) velocity deficit with respect to $M_\mathrm{prec}$ and velocity magnitude at the sides of the wake normalized with the velocity magnitude obtained in the precursor domain at hub height. All quantities are predicted by the fitting model shown in Equation~\ref{eq:fitting_model}. Moreover, the $x$-axis is rescaled with $x_0=30$ km which denotes the beginning of the wake region.}
	\label{fig:fit_wake_properties_dh}
\end{figure}

Next, we discuss the differences in wake width, strength and deflection within the wake and at its sides by using the fitting model reported in Equation \ref{eq:fitting_model}. Since the capping-inversion height of case H300-$\Delta$H500 is in between the one of case H300 and H500, we also include the latter case in these comparisons. Although the wake width at the beginning of the wake region is similar among the three cases, we can see in Figure \ref{fig:fit_wake_properties_dh}(a) that the wake narrowing is more accentuated in case H300-$\Delta$H500. Further, Figure \ref{fig:fit_wake_properties_dh}(b) illustrates that case H300-$\Delta$H500 has a very minor wake deflection. This means that the two opposite effects triggered by the Coriolis force which are responsible for a clockwise or anticlockwise flow rotation of the farm wake are in balance for this atmospheric state. The velocity deficit with respect to $M_\mathrm{side}$ is shown in Figure \ref{fig:fit_wake_properties_dh}(c). The recovery rate of case H300-$\Delta$H500 is very similar to the one of cases H300 and H500. However, $\Delta M$ is lower than case H300, as already observed in Figure \ref{fig:xz_slices_dh}(b). Finally, the velocity at the wake sides is shown in Figure \ref{fig:fit_wake_properties_dh}(d). A ticker capping inversion reduces the flow-blockage effect, so that the flow acceleration at the wake sides is also reduced significantly. This explains the very similar trend observed for the $M_\mathrm{side}$ profile of cases H300-$\Delta$H500 and H500.

\begin{figure}
	\centering
	\includegraphics[width=0.55\textwidth]{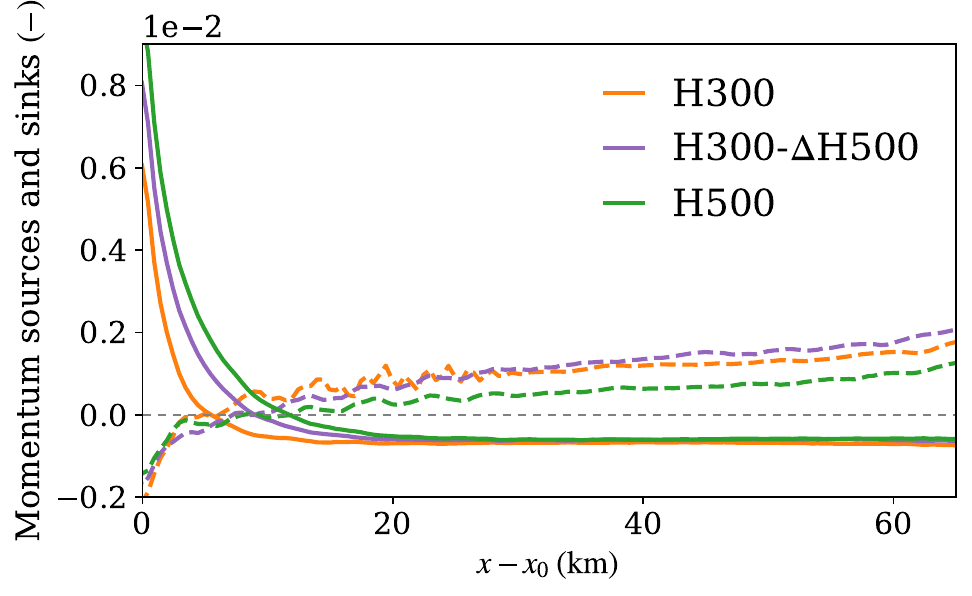} 
	
	\caption{Streamwise variation of $\Delta \mathcal{F}_{uw,z}$ (full line) and $\Delta \mathcal{A}_{uv,y}$ (dashed line) in the turbine region normalized by the time averaged total wind-farm thrust force of the respective case and the control volume width. We note that the $x$-axis is rescaled with $x_0=30$ km which denotes the beginning of the wake region.}
	\label{fig:x_mom_comparison_dh}
\end{figure}

Finally, we compare the two dominant terms that contribute to wake recovery, that is $\Delta \mathcal{F}_{uw,z}$ and $\Delta \mathcal{A}_{uv,y}$, between cases H300, H300-$\Delta$H500 and H500. The results are shown in Figure \ref{fig:x_mom_comparison_dh}. The vertical turbulent entrainment of momentum in case H300-$\Delta$H500 shows a slower decay than in case H300 along the streamwise direction. However, its magnitude is still lower than in case H500. The mean flow entrainment along the spanwise direction is twice as strong than in case H500, and even higher than in case H300, particularly in the far-wake region. The combination of these two factors makes the wake recovery rate in case H300-$\Delta$H500 faster than in case H300, but still slower with respect to case H500.

\section{Mass budget analysis}\label{app:mass_balance}
In this appendix, we investigate the mass balance in the farm wake. Hence, we first take a time average of the continuity equation, further integrating it over the control volume $\Omega$ to average out local oscillations -- see Section~\ref{sec:momentum_analysis} for more information on $\Omega$. Additionally, we apply the divergence theorem, which enables us to eliminate the divergence operator and transition from a volume to a surface integral. As a result, the mass budget equation reads as
\begin{equation}
	\underbrace{-\biggl[ \int_{\Gamma_x} \bar{u} d\Gamma_x\biggl]_{x_1}^{x_2}}_{\Delta\mathcal{A}_{u,x}} 
	\underbrace{-\biggl[ \int_{\Gamma_y} \bar{v} d\Gamma_y\biggl]_{y_1}^{y_2}}_{\Delta\mathcal{A}_{v,y}} 
	\underbrace{-\biggl[ \int_{\Gamma_z} \bar{w} d\Gamma_z\biggl]_{z_1}^{z_2}}_{\Delta\mathcal{A}_{w,z}} = 0
	\label{eq:mass}
\end{equation}
where the terms $\Delta\mathcal{A}_{u,x}$, $\Delta\mathcal{A}_{v,y}$ and $\Delta\mathcal{A}_{w,z}$ represent the difference in streamwise, spanwise and vertical mass flux, respectively. We note that, due to the sign convention chosen, these terms are positive when the inflow mass flux is higher than the outflow and negative when the opposite occurs. For example, the term $\Delta\mathcal{A}_{u,x}$ is positive when the streamwise mass flux across the surface $\Gamma_{x_1}$ is higher than the streamwise mass outflow through $\Gamma_{x_2}$. This means that an increase in streamwise velocity along the $x$ direction causes this term to be negative. 

\begin{figure}
	\includegraphics[width=1\textwidth]{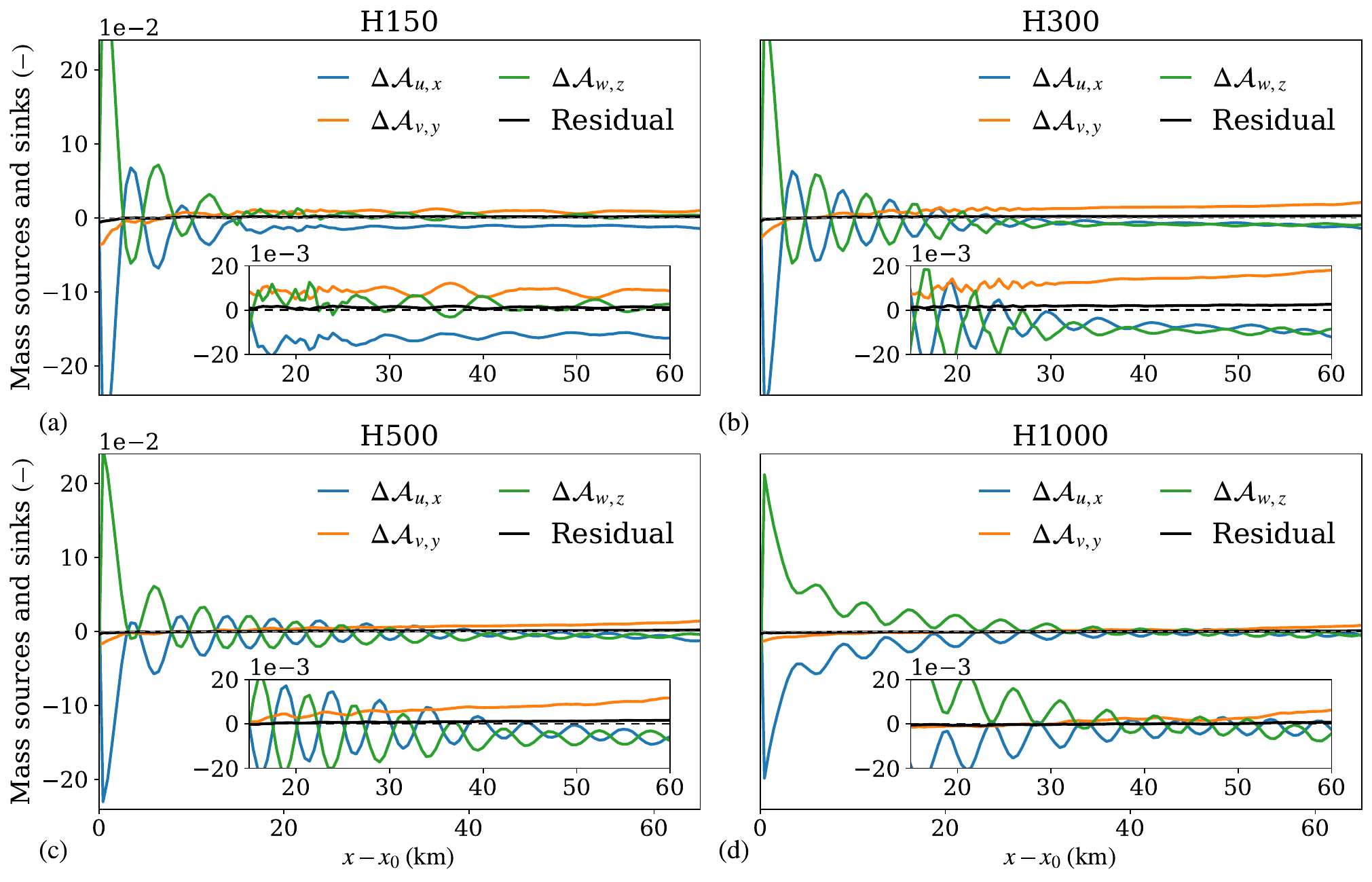} 
	
	\caption{Streamwise variation of mass sources and sinks for cases (a) H150, (b) H300, (c) H500 and (d) H1000 normalized by the time averaged total wind-farm thrust force times the geostrophic wind and the control volume width. We note that only the region downwind of the farm is considered. Moreover, the $x$-axis is rescaled with $x_0=30$ km which denotes the beginning of the wake region.}
	\label{fig:mass_balance}
\end{figure}

Figure \ref{fig:mass_balance} shows the streamwise evolution of all terms of Equation \ref{eq:x_mom} for all cases. First, we notice that the terms $\Delta\mathcal{A}_{u,x}$ and $\Delta\mathcal{A}_{w,z}$ are out of phase. This results from the fact that a flow acceleration (negative $\Delta\mathcal{A}_{u,x}$) causes the flow to move downward (positive $\Delta\mathcal{A}_{w,z}$), and vice versa. The strong oscillatory behaviour is due to the presence of trapped waves. The oscillation frequency is equal in all cases since the gravity-wave horizontal wavelength does not depend on $H$. The latter measures about $4.5$~km (see \cite{Vosper2004,Sachsperger2015}), which is in line with the period of the oscillations shown in Figure \ref{fig:mass_balance}. In case H150, we observe limited mass fluxes along the vertical direction compared to other cases. This is due to the close vicinity of the capping inversion, which severely limits vertical motion. Further downstream, $\Delta\mathcal{A}_{u,x}$ and $\Delta\mathcal{A}_{u,y}$ assume a similar magnitude, meaning that the mass sink generated by the accelerating flow along the streamwise direction is solely compensated by spanwise entrainment. A different behaviour is shown in Figure \ref{fig:mass_balance}(b), where results for case H300 are shown. The higher capping inversion accounts for a higher mass transfer along the vertical direction, although the latter remains small if compared to cases with higher $H$. Therefore, we observe a stronger entrainment along the spanwise direction, which compensate for the negative $\Delta\mathcal{A}_{u,x}$ and $\Delta\mathcal{A}_{w,z}$ terms. A similar behaviour is observed for case H500, although $\Delta\mathcal{A}_{v,y}$ is roughly half of the one attained in the H300 case. Spanwise entrainment becomes negligible in the H1000 case, as illustrated in Figure \ref{fig:mass_balance}(d). Here, the wake recovery is facilitated by mass fluxes along the vertical direction. Finally, we remark that the vertical mass flux along the $\Gamma_z$ surface located at $z_1$ is negligible in all cases (not shown).

\end{document}